\documentclass[12pt,letterpaper]{JHEP3} \usepackage{cite}
\usepackage{epsfig}

\newcommand{\solar}{\odot}

\title{Predicting the Cosmological Constant from the Causal Entropic
  Principle}

\author{%
Raphael Bousso$^a$, Roni Harnik$^b$, 
Graham D. Kribs$^c$, and Gilad Perez$^d$\\
\llap{$^a$}
Center for Theoretical Physics, Department of Physics, 
University of California,\\ 
Berkeley, CA 94720-7300, U.S.A. {\em and} \\
Lawrence Berkeley National Laboratory, 
Berkeley, CA 94720-8162, U.S.A. 
\vspace*{2mm} \\ 
\llap{$^b$}
Stanford Linear Accelerator Center, Stanford University, \\
Stanford, CA 94309, U.S.A. {\em and} \\
Physics Department, Stanford University, Stanford, CA 94305, U.S.A.
\vspace*{2mm} \\
\llap{$^c$}
Department of Physics and Institute of Theoretical Science, 
University of Oregon, \\ Eugene, OR 97403, U.S.A. 
\vspace*{2mm} \\
\llap{$^d$}
C.N. Yang Institute for Theoretical Physics, State University of New York, \\
Stony Brook, NY 11794-3840, U.S.A.}

\abstract{We compute the expected value of the cosmological constant
  in our universe from the Causal Entropic Principle.  Since observers
  must obey the laws of thermodynamics and causality, the principle
  asserts that physical parameters are most likely to be found in the
  range of values for which the total entropy production within a
  causally connected region is maximized.  Despite the absence of more
  explicit anthropic criteria, the resulting probability distribution
  turns out to be in excellent agreement with observation.  In
  particular, we find that dust heated by stars dominates the entropy
  production, demonstrating the remarkable power of this thermodynamic
  selection criterion.  The alternative approach---weighting by the
  number of ``observers per baryon''---is less well-defined, requires
  problematic assumptions about the nature of observers, and yet
  prefers values larger than present experimental bounds.}

\preprint{\hepth{0702115}}

\begin{document}

\section{Introduction}
\label{sec-intro}

The discovery that the universe is in a period of accelerated
expansion~\cite{Per98,Rie98}, combined with an accurate accounting of
the total, matter, and radiation components of the energy density
\cite{Spergel:2006hy}, provide overwhelming evidence for dark energy.
These measurements are completely consistent with the interpretation
of dark energy as a non-zero cosmological constant, $\Lambda$.  This
has undermined the hope that the energy of our vacuum is uniquely
determined by fundamental theory.  Instead, it lends credence to the
hypothesis that the cosmological constant is an environmental
variable, which takes on different values in widely separated regions
of the universe.

The observed vacuum energy density\footnote{Unless indicated
  otherwise, all observed values in this paper are taken from
  Ref.~\cite{Tegmark:2005dy}.  Where no explicit units are given, we set
  $\hbar=G_{\rm N}=c=k_{\rm B}=1$.}
\begin{equation}
\rho_\Lambda=\frac{\Lambda}{8\pi} = (1.25\pm 0.25)\times 10^{-123}~,
\label{eq-cc}
\end{equation} 
is at least $55$ orders of magnitude smaller than what would be
expected from the standard model of particle physics (see, e.g.,
Ref.~\cite{Polchinski:2006gy} for a recent review).  The environmental
approach does not assert that this tiny value is inevitable, or even
typical among all possible values.  Rather, it aims to show that it is
not atypical among values measured by observers.

A number of conditions must be satisfied for the environmental
approach to work.  The first, obviously, is that fundamental theory
must {\em admit}\/ the observed value of the vacuum energy.  This can
happen without explicit tuning if the theory gives rise to an enormous
number $N$ of different vacua.  Of course, typical values will be of
order unity, but if they are randomly distributed, they can form a
dense spectrum, or ``discretuum'', with average spacing of order
$1/N$.  If $N\gg 10^{123}$, then it is likely that the observed value
is included in the spectrum.  Thus, the approach really depends on
whether fundamental theory (which, presumably, is more or less unique)
admits a sufficiently dense discretuum.

The second condition is that the observed value must be dynamically
{\em attainable\/}, starting from generic initial conditions.  With
$N\gg 10^{123}$ possibilities, there is no reason for the universe to
start out in a vacuum like ours.  The environmental approach therefore
depends on a means to start from some generic initial value and later
realize the observed value, either as a branch in the wavefunction or
as a particular spacetime region embedded in a vast universe.

Finally, the environmental approach requires an explanation of why we
happen to {\em observe\/} such an unusually small vacuum energy.  Most
values of $\rho_\Lambda$ in the discretuum will be of order unity, and
only a fraction (in the simplest case, a fraction of order
$10^{-123}$) will have a magnitude as small as the observed value.  It
is not enough to show that the small value given by Eq.~(\ref{eq-cc})
is possible; one must also show that it is {\em not unlikely\/} to be
observed.

The first condition appears to be satisfied by string theory, which
admits as many as $10^{500}$ long-lived metastable
vacua~\cite{BP,KKLT,DenDou04b,Sil01,DouKac06} (see Ref.~\cite{Sch06}
for a discussion of earlier work).  The second condition can then be
met because the vacua are metastable and can decay into one
another.\footnote{In general, long-lived metastability implies that
  all matter is diluted before the next decay occurs, so the mechanism
  depends on efficient reheating in the new region.  This rules out
  models that reduce the cosmological constant
  gradually~\cite{BT1,BT2}. In the string landscape, the vacuum
  preceding ours was likely to have had an enormous cosmological
  constant.  Its decay acted like a big bang for the observed universe
  and allowed for efficient reheating~\cite{BP}.} 

In this paper we address the third condition.  We will use a novel
approach, the Causal Entropic Principle, to argue that the observed
value of $\rho_\Lambda$ is not unlikely.  Our main result is shown in
Fig.~\ref{fig-showmethemoney}.
\EPSFIGURE[!t]{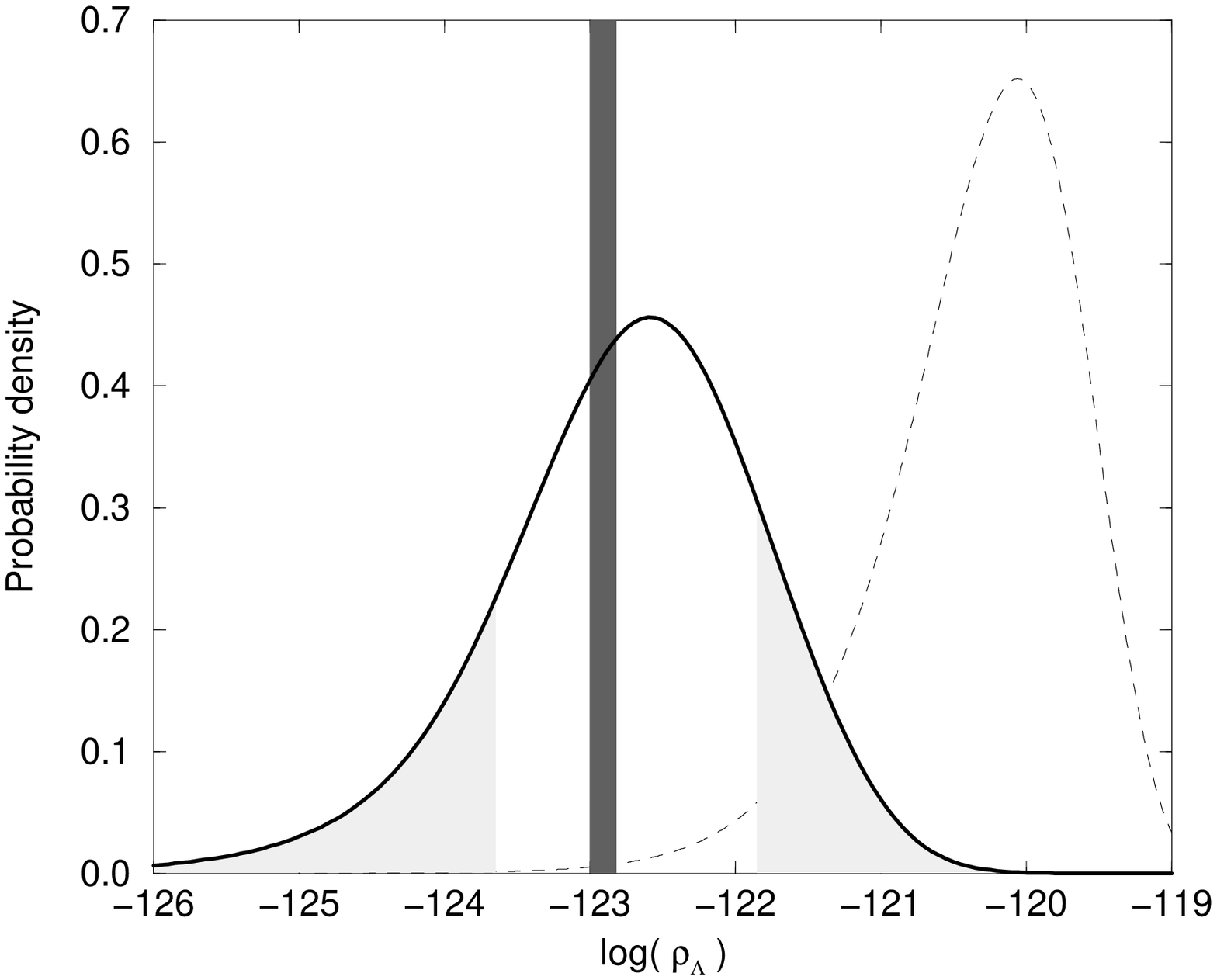,width=.8\textwidth}{\label{fig-showmethemoney}
  The probability distribution of the vacuum energy measured by
  typical observers, computed from the Causal Entropic Principle, is
  shown as a solid curve.  The values consistent with present
  cosmological data, in the shaded vertical bar, are well inside the
  $1\sigma$ region (shown in white), and hence, not atypical.  For
  comparison, the dashed line shows the distribution derived by
  estimating the number of observers per baryon.  Unlike our curve, it
  assumes that galaxies are necessary for observers; yet, the observed
  value is very unlikely under this distribution.  For more details
  about both curves, see Figures 2 and 8.}

The Causal Entropic Principle is based on two ideas: any act of
observation increases the entropy, and spacetime regions that are
causally inaccessible should be disregarded.  It assumes that on
average, the number of observations will be proportional to the amount
of matter entropy produced in a causally connected region, $\Delta S$.
Vacua should be weighted by this factor to account for the rate at
which they will be observed.

Crucially, the size of the causal diamond is inversely proportional to
the vacuum energy, so smaller values of $\rho_\Lambda $ allow for
greater complexity.  This compensates for the scarcity of vacua with
small $\rho_\Lambda $.  As a result, $\rho_\Lambda$ prefers to take a
value such that vacuum energy begins to dominate near the time of peak
entropy production.  

We will find that entropy production in the causal diamond is
dominated by dust particles heated by stars.  This is an important
result in its own right: our weight, $\Delta S$, is a simple physical
quantity that turns out to be sensitive to the existence of galaxies,
stars, and heavy elements.

We will show that the entropy production rate peaked approximately 2
to 3.5 billion years after the big bang.  It is this time-scale,
rather than the time of galaxy formation, which governs our prediction
of the cosmological constant, and it prefers a range of $\rho_\Lambda
$ that is in very good agreement with observation.

The same result also explains the so-called coincidence problem or
``why now'' problem.  According to the Causal Entropic Principle,
typical observers will exist when most of the entropy production in
the causal diamond occurs.  Our result ensures that this happens
during the era when the matter and vacuum energy densities are
comparable.

\paragraph{Outline}

Historically, discussions of the third condition---why do we observe
an unusually small $\rho_\Lambda$---have focussed on anthropic
selection effects, which we discuss in Sec.~\ref{sec-weighting}.  Long
before the discovery of the string landscape, it was noted that not
all values of $\rho_\Lambda$ are compatible with the existence of
observers~\cite{DavUn,Linde84,Sak84,Ban85,BarTip,Linde}.  This culminated in Weinberg's
successful prediction~\cite{Wei87} that a small non-zero value of
$\rho_\Lambda$ would be observed, which we review in
Sec.~\ref{sec-weinberg}.  Weinberg's assumption was relatively modest:
Observers require galaxies.  But astronomers have since discovered
galaxies that would have formed even if $\rho_\Lambda$ had been more
than a thousand times larger than the observed value.  This leaves a
large discrepancy between theory and observation.

A possible resolution is to ask not only about the existence of
observers, but to weight vacua by the {\em number\/} of observers they
contain.  This number is generically infinite or zero, so a
regularization scheme is needed.  A popular approach is to weight by
the number of observers per baryon.  In Sec.~\ref{sec-opb}, we argue
that this approach is both poorly motivated and, in a realistic
landscape, poorly defined.  Moreover, it does not resolve the conflict
with observation.  To mitigate the discrepancy, one is forced to posit
increasingly specific conditions for life, such as the chemical
elements required.  Indeed, to do reasonably well, one must suppose
that observers can only arise in galaxies as large as ours---a very
strong assumption, for which there appears to be no evidence.

In Sec.~\ref{sec-cep}, we motivate and discuss a different approach to
this problem.  The Causal Entropic Principle weights each vacuum by
the amount of entropy, $\Delta S$, produced in a causally connected
region~\cite{Bou06}.  This is the largest spacetime region that can be
probed and across which matter can interact.  Since observation
requires free energy, it is natural to expect that the number of
observers will scale, on average, with $\Delta S$.  In other words, we
demand nothing more than that observers obey the laws of
thermodynamics.  This is far weaker even than Weinberg's criterion,
and one might be concerned that it will not be sufficiently
restrictive.  Yet, in combination with the causal cutoff, this minimal
requirement becomes a powerful predictive tool.

This is demonstrated in Sec.~\ref{sec-calculation}, where we use the
Causal Entropic Principle to derive the probability distribution for
$\rho_\Lambda$ over universes otherwise identical to ours.  We begin,
in Sec.~\ref{sec-metric}, by computing the geometry of the causal
diamond for general $\rho_\Lambda $.  We are mainly interested in its
comoving volume as a function of time, $V_{\rm c} (t)$.  In
Sec.~\ref{sec-entropy}, we consider important mechanisms by which
entropy is produced within our causal diamond.  We estimate
contributions from stars, quasars, supernovae, and other processes.
We find that the leading contribution to $\Delta S$ comes from
infrared photons emitted by interstellar dust heated by stars.

In Sec.~\ref{sec-rate}, we analyze this leading contribution in more
detail.  We compute $d S/(d V_{\rm c} \, dt)$, the rate at which
entropy is produced per unit comoving volume per unit time.  This rate
will depend on $\rho_\Lambda$ because large $\rho_\Lambda$ disrupts
galaxy formation and thus star formation.  Interestingly, however, it
turns out that this dependence is not important for our final result.
In Sec.~\ref{sec-total}, we integrate the rate found in
Sec.~\ref{sec-rate} over the causal diamond determined in
Sec.~\ref{sec-metric}.  This yields the weight factor, $\Delta
S(\rho_\Lambda )$.  We display the resulting probability distribution
for $\rho_\Lambda$, and we note that the observed vacuum energy lies
in the most favored range.

In Sec.~\ref{sec-discussion}, we pinpoint the origins and discuss some
implications of our main findings.  We also identify important
intermediate results.

\paragraph{Extensions}

In the interest of time and clarity, we have limited our task.  We use
the Causal Entropic Principle solely to compute a probability
distribution over positive values of the vacuum energy, holding all
other physical parameters fixed.  This is the case most frequently
studied in the literature, making it straightforward to compare our
result with those obtained from the traditional method of weighting by
observers-per-baryon~\cite{MarSha97,GarLiv99,Efs95,Vil04}.

In other words, our distribution is conditioned on the assumptions
that $\rho_\Lambda>0$, and that all low energy physics is the same as
in our vacuum.  We ask only about the probability distribution of
$\rho_\Lambda$ in this subspace of the landscape.  This is a valid
consistency check: Suppose that the observed value were disfavored on
a subspace picked out by other observed parameters.  Then it would
only become less likely when the distribution is extended over the
entire landscape, and so the model would conflict with observation.

Because negative values of $\rho_\Lambda$ are tightly constrained by
standard (though questionable) anthropic arguments, the main challenge
for environmental approaches has been to suppress large positive
values of $\rho_\Lambda$.  For this purpose, it suffices to
concentrate on the subset of vacua with $\rho_\Lambda>0$.  This
simplifies our analysis, since negative $\rho_\Lambda$ lead to a
different class of metrics.  (Interestingly, a preliminary analysis
indicates that negative values will be somewhat favored by the Causal
Entropic Principle, though not by enough to render the observed value
unlikely.)

It will be interesting to use the Causal Entropic Principle to compute
the probability distribution of other parameters, such as the
amplitude of primordial density perturbations, $\delta\rho/\rho$, the
spatial curvature, $k$, or the baryon to photon ratio, $\eta$.  A
crucial task will be to estimate the probability distribution over
multiple parameters at once, since this is a much more stringent test
for the environmental approach.  For example, consider a distribution
over two parameters, $\rho_\Lambda$ and $\delta\rho/\rho$.  When
weighting by observers-per-baryon, the upper bound on $\rho_\Lambda$
arises from the requirement that galaxies can form.  Hence, it would
seem highly favorable to increase $\delta\rho/\rho$, since
$\rho_\Lambda$ could then be increased by the third power of the same
factor.  This would render the observed values of both $\rho_\Lambda$
and $\delta\rho/\rho$ exceedingly unlikely.  As we will discuss in
Sec.~\ref{sec-understand}, however, galaxy formation is not a
significant constraint on $\rho_\Lambda$ in our approach.  Hence, we
expect this problem to be virtually absent.

Given a multivariate distribution, one can ask about the probability
distribution over one parameter (say, $\log\rho_\Lambda$) with other
parameters integrated out.  As more parameters are allowed to vary,
the distribution for $\log\rho_\Lambda$ is likely to become broader
after they are integrated out.  Yet, the observed value must remain
typical if the environmental approach is to succeed.  The most radical
choice is to study the distribution of $\log\rho_\Lambda$ after
integrating out all other parameters characterizing the landscape.
This would be tantamount to deriving the typical range of $\log
\rho_\Lambda $ from fundamental theory alone, without reference to
parameters specific to our own vacuum.  This would have been
impossible in conventional approaches.  But as we discuss in
Sec.~\ref{sec-deltas}, $\Delta S$ may depend simply on $\rho_\Lambda $
when averaged over the entire landscape.  Hence, the Causal Entropic
Principle puts this task within our reach.

\section{Approaches to weighting vacua}
\label{sec-weighting}

\subsection{The Weinberg bound}
\label{sec-weinberg}

Weinberg~\cite{Wei87} estimated the range of $\rho_\Lambda$ compatible
with galaxy formation.  No galaxies form in regions where
$\rho_\Lambda$ exceeds the matter density $\rho_{\rm m}$ at the time
when the corresponding density perturbations become nonlinear
(assuming otherwise identical physics).  If we grant that galaxies are
a prerequisite for the existence of observers, then these regions will
not contain observers, and such values of $\rho_\Lambda$ will not be
observed.  Combined with a similar argument\footnote{With
  $\rho_\Lambda<0$, the universe will recollapse after a time of order
  $|\rho_\Lambda|^{-1/2}$.  If one assumes that most observers emerge
  only after several billion years, then an upper bound on
  $(-\rho_\Lambda)$ results by requiring that the universe should not
  recollapse too soon~\cite{BarTip}.---We will consider only positive
  values of $\Lambda$ in this paper.} for negative values of
$\rho_\Lambda$, Weinberg~\cite{Wei87} found that only values in the
interval
\begin{equation}
-1<\frac{\rho_\Lambda}{\rho_{\rm m}}<550
\label{eq-littlesteve}
\end{equation}
will be observed.  

The upper bound is larger than 1 because the matter density today,
$\rho_{\rm m}$, has been diluted since galaxy formation, by the
redshift factor $(1+z_{\rm gal})^3$.  Weinberg used $z_{\rm
  gal}\approx 4.5$, but in the meantime, dwarf galaxies have been
discovered at redshifts as high as $z_{\rm gal}\approx
10$~\cite{Loeb:2006en}, raising the upper bound:
\begin{equation}
-1<\frac{\rho_\Lambda}{\rho_{\rm m}}<5000~.
\label{eq-bigsteve}
\end{equation}

The observed ratio of vacuum to matter energy density is much smaller
than the upper bound:
\begin{equation}
\frac{\rho_\Lambda}{\rho_{\rm m}}\approx 2.3~.
\end{equation}
In other words, it would appear that the observed value of
$\rho_\Lambda$ is in fact quite unlikely, even allowing for the
anthropic constraint, Eq.~(\ref{eq-bigsteve}).

Let us rephrase Weinberg's argument in a more general language.  The
probability distribution for observed values of $\rho_\Lambda$ can be
written as
\begin{equation}
  \frac{dP}{d\rho_\Lambda}\propto 
  w(\rho_\Lambda) \frac{dp}{dN} \frac{dN}{d \rho_\Lambda}~,
\end{equation}
where $N$ is the number of vacua with vacuum energy smaller than
$\rho_\Lambda$, and $p$ is the total prior probability for these
vacua.  Since $\rho_\Lambda=0$ is not a special point, vacua in the
landscape are uniformly distributed when averaged over intervals of
$\rho_\Lambda$ of order $10^{-123}$ or smaller near $\rho_\Lambda=0$:
\begin{equation}
  \frac{dN}{d \rho_\Lambda}=\mbox{const}~.
\end{equation}
Before anthropic selection, it is reasonable to assume that all vacua
are equally likely:\footnote{Dynamical effects can modify this flat
  prior~\cite{SchVil06}.  We shall assume that the resulting
  distribution remains effectively flat, at least for small
  $|\rho_\Lambda|$.  Models violating this assumption are unlikely to
  solve the cosmological constant problem.}
\begin{equation}
\frac{dp}{dN}=\mbox{const}~.
\end{equation}
The anthropic condition of galaxy formation assigns a weight $w=1$ to
vacua in the range of Eq.~(\ref{eq-bigsteve}), and $w=0$ to all other
vacua.  Thus, all values of $\rho_\Lambda$ in the interval of
Eq.~(\ref{eq-bigsteve}) are equally likely,
$dP/d\rho_\Lambda=\mbox{const}$.

Restricting to vacua with $\rho_\Lambda>0$, it is instructive to
consider the probability distribution as a function of $\log
\rho_\Lambda$,
\begin{equation}
  \frac{dP}{d\log\rho_\Lambda} = \rho_\Lambda\frac{dP}{d\rho_\Lambda}
  \propto \rho_\Lambda w(\rho_\Lambda) ~.
\label{eq-plog}
\end{equation} 
With the above, ``binary'' weight, this distribution will be a growing
exponential of $\log \rho_\Lambda$, with a sharp cutoff at $\log
\rho_\Lambda\approx -120$ from the upper bound in
Eq.~(\ref{eq-bigsteve}).  The observed value, $\log
\rho_\Lambda\approx -123$, is suppressed by about 3 orders of
magnitude compared to values near the cutoff.

\subsection{Weighting by observers per baryon}
\label{sec-opb}

In order to reduce this discrepancy, one would need to go beyond the
binary question of whether or not there are observers.  Surely the
{\em number\/} of observers will depend on the cosmological constant
and will begin to decrease for values of $\rho_\Lambda$ smaller than the 
upper bound in Eq.~(\ref{eq-bigsteve}).  If we weight vacua
by this number, perhaps this will be more effective at suppressing
large values of $\rho_\Lambda$ than a simple binary filter.

Unfortunately, this strategy is not well-defined without a
regularization scheme.  The dynamics of eternal inflation results in a
multiverse containing an infinite number of regions for every value of
$\rho_\Lambda$.  Each region is an open universe with infinite spatial
volume at all times.  (The hyperbolic spatial geometry reflects the
symmetries preserved by a vacuum bubble formed in a first-order phase
transition from a higher metastable vacuum.)  If a vacuum admits any
observers at all, their number will be infinite.

Various authors~\cite{MarSha97,GarLiv99,Efs95} have proposed to weight
vacua by the number of observers per baryon, or per photon, or per
unit matter mass.  But none of these choices are particularly well
motivated.  If there are infinitely many baryons, why should it matter
how efficiently they are converted to observers?  Why is a vacuum less
likely to be observed if a smaller fraction of its mass becomes
observers, as long as there are infinitely many of them?  

More importantly, these regularization methods are not universally
defined.  This makes them inapplicable in a rich landscape, where we
will eventually be forced to consider vacua with very different low
energy physics.  Two vacua may have different baryon-to-photon ratios,
so that the above weighting methods are inequivalent; which should we
choose?  Indeed, it seems unlikely that a standard definition of
``baryon'' can be given that would be meaningful in all
vacua.\footnote{Ref.~\cite{Vilenkin} proposes an interesting method
  for defining a unit comoving volume in different vacua, in the limit where 
  bubbles preserve an exact $SO(3,1)$ symmetry.}

These difficulties arise because the regularization refers to a
reference particle species such as ``baryons''.  But it also refers to
``observers'', and this leads to additional problems.  It seems
virtually impossible to define what an observer is in vacua with
different low energy physics.  Even in our own universe, it is unclear
how to estimate the number of observers per baryon.  One approximation
is to assume that it will be proportional to the fraction of baryons
that end up in galaxies.  But this fraction depends strongly on the
minimum mass of a galaxy capable of harboring observers, $M_*$, which
is not known.

\EPSFIGURE[t]{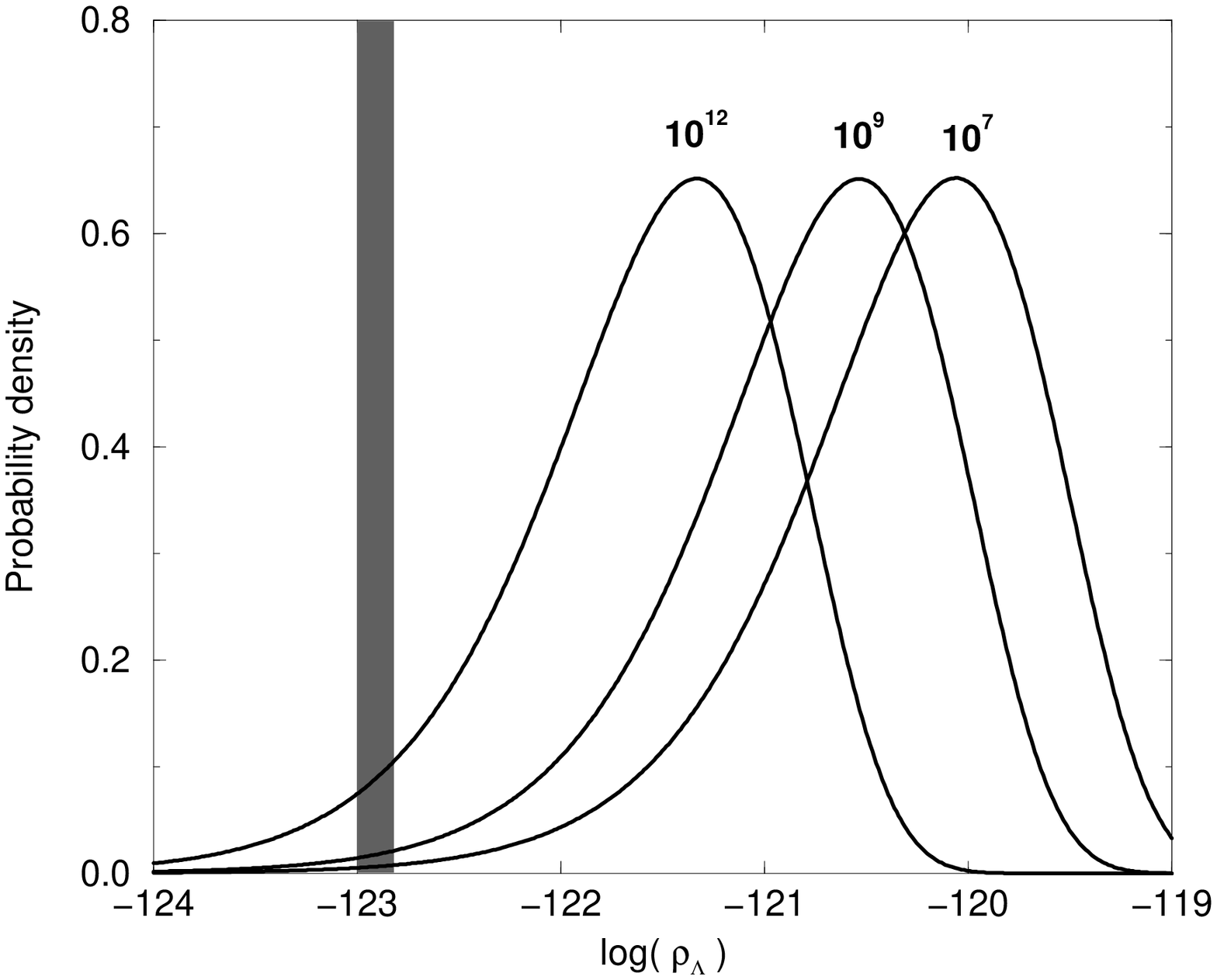,width=.7\textwidth}{\label{anthrocurves}
  Weighted by ``observers per baryon'', the probability distribution
  for $\rho_\Lambda$ depends strongly on specific assumptions about
  conditions necessary for life.  Three curves are shown,
  corresponding to different choices for the minimum required mass of
  a galaxy: $M_* = (10^7, 10^9, 10^{12}) M_\odot$.  In neither case is
  the observed value (vertical bar) in the preferred range.  The
  choice $M_*=10^7M_\odot$ (also shown in Fig.~1) corresponds to the
  smallest observed galaxies. The choice $M_* = 10^{12} M_\odot$
  minimizes the discrepancy with observation but amounts to assuming
  that only the largest galaxies can host observers.  By contrast, the
  Causal Entropic Principle does not assume that observers require
  structure formation, let alone galaxies of a certain mass; yet its
  prediction is in excellent agreement with the observed value (see
  Fig.~8).}
Figure~\ref{anthrocurves} shows probability distributions for
$\rho_\Lambda$, under various assumptions for $M_*$.  Dwarf galaxies
as small as $M_*\sim 10^7 M_\odot$ have been detected.  With this
choice, the observed value of $\Lambda$ is nearly three orders of
magnitude, or 3.5$\sigma$, below the median value.

There is some evidence that galaxies with mass below $10^9 M_\odot$
will not retain the heavy elements produced in supernova
explosions~\cite{Vil04}.  Under the {\em additional assumption\/} that
such elements are required for life, one may then set $M_*$ to this
larger value.  But the resulting prediction remains unsatisfactory.
The median exceeds the observed value by a factor of more than two
orders of magnitude, or 2.9$\sigma$.

One can speculate that for some reason, life requires a galaxy as
large as our own, or perhaps even a larger group~\cite{MarSha97,Vil04}
($M_*=10^{12} M_\odot$).  Then the observed value is about 
$1.8\sigma$, or a factor of $22$, below the median of the
predicted distribution.  However, at present we can see no compelling
arguments for this extreme choice.  Thus, the weighting by
observers-per-baryon leads to a dilemma: Either it requires overly
specific and questionable assumptions, or else it faces a real
conflict with the data.

As these difficulties demonstrate, weighting by observers-per-baryon
may not be the correct way to compute probabilities in the landscape.
We will now argue for a different approach, which is always
well-defined.  It will allow us to assume nothing more about observers
than that they respect the laws of thermodynamics.

\subsection{Weighting by entropy production in the causal diamond}
\label{sec-cep}

\paragraph{Causal Entropic Principle}
In this paper we will compute the probability distribution for
$\rho_\Lambda$ based on the Causal Entropic Principle, which is
defined by the following two conjectures~\cite{Bou06}:
\begin{itemize} 
\item[(1)]{The universe consists of one causally connected region, or
    ``causal diamond''.  Larger regions cannot be probed and should
    not be considered part of the semi-classical geometry.}
\item[(2)]{The number of observations is proportional to $\Delta S$,
  the total entropy production in the causal diamond.}
\end{itemize}
\EPSFIGURE[t]{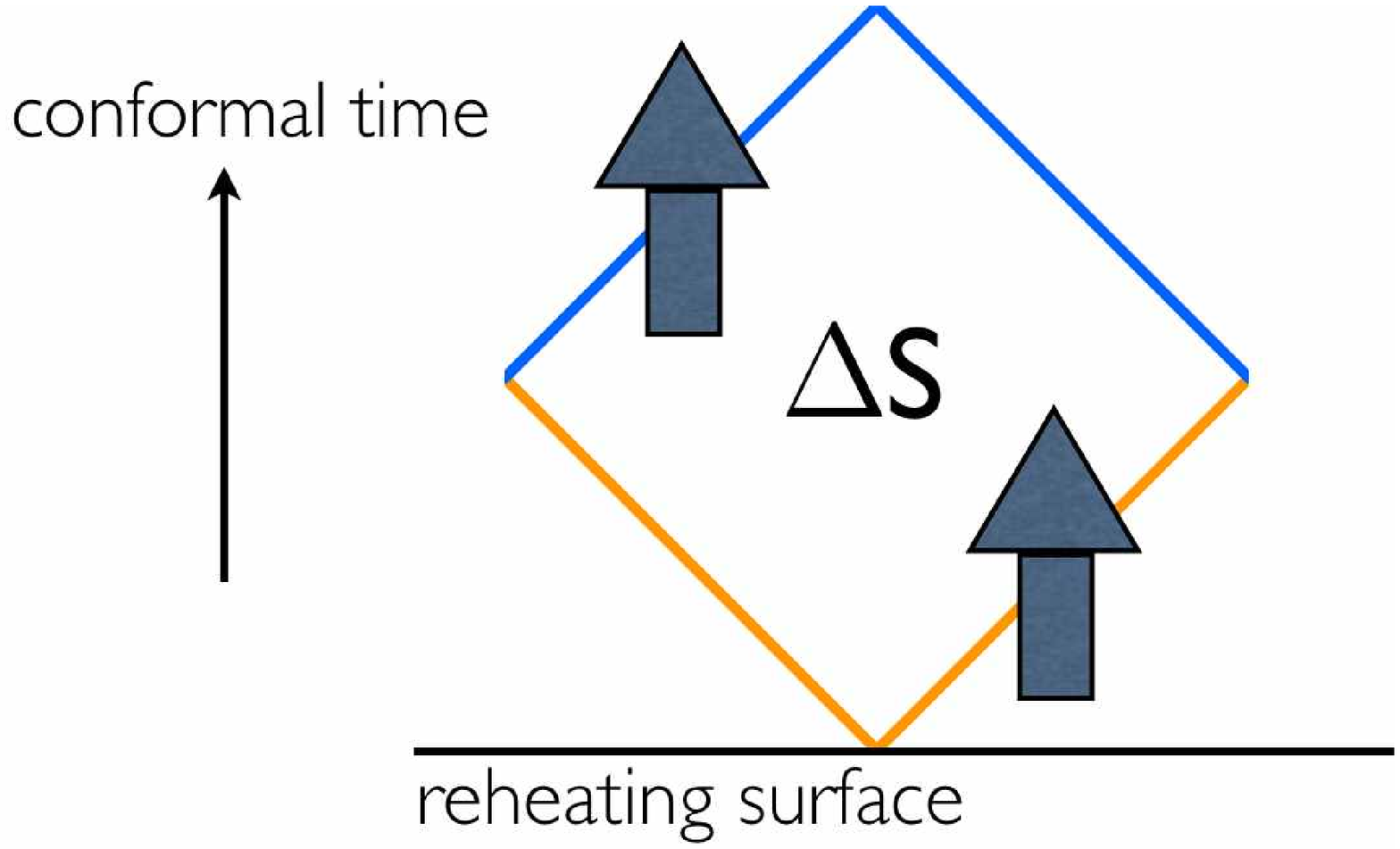,width=.6\textwidth}{\label{fig-deltas} A
  causal diamond is the largest spacetime region over which matter can
  interact.  It is delimited by a future lightcone from a point on the
  reheating surface (orange/light), and by a past lightcone from a late-time
  event (blue/dark); in the case of de~Sitter vacua this is the
  cosmological horizon.  A vacuum should be weighted by the number of
  observations made in this spacetime region.  Since observation
  requires free energy, we expect that on average, this number will be
  proportional to the amount of entropy, $\Delta S$, produced in the
  causal diamond.  Entropy entering through the bottom cone (bottom
  arrow), such as the CMB, does not contribute to this entropy
  difference.}

Before motivating the two conjectures, let us first clarify the key
terms---causal diamond and entropy---they refer to.

The causal diamond~\cite{Bou00a} is the largest region of spacetime
causally accessible to a single observer.  It is defined by the
intersection of the past light-cone of a late-time point on the
worldline with a future light-cone of an early-time point, shown in
Fig.~\ref{fig-deltas}.  We choose this time to be the time of
reheating, since no matter existed before then.  (In vacua with no
reheating, $\Delta S$ vanishes independently of the choice of the
causal diamond.  This case does not arise here, since we are holding
all parameters other than $\rho_\Lambda$ fixed.)  Only after reheating
can matter begin to interact and commence the formation of complex
structures, at most at the speed of light.

In a vacuum with negative cosmological constant, the tip of the top
cone would be on the big crunch.  In any metastable vacuum with
positive cosmological constant, like ours, the top cone is given by
the de~Sitter event horizon.

One may question the universality of ``reheating surface'' or our use
of an event horizon (a global concept) in a vacuum with finite
lifetime, so let us give a more careful definition.  A causal diamond
can be triggered (that is, a bottom cone drawn) as soon as there is
entropy in the matter sector.  Reheating is a special case: during
inflation, all entropy is in the gravitational sector (the growing
horizon of the inflationary universe), but reheating generates matter
entropy (mostly radiation, which we include in this class).  After
vacuum domination in a metastable de~Sitter region, the diamond
empties out at an exponential rate.  If we enlarge the diamond by
choosing a later point for the tip, and the additional spacetime
volume contains no matter (which will be generic at late times), there
is no point in going further.  The total amount of matter enclosed by
the top cone will be the same as if the vacuum had been completely
stable.  Once the metastable vacuum decays, reheating may again occur,
in which case a new diamond is triggered.  This definition implies
that if a decay or phase transition happens while there is still
matter around, the two vacua should not be considered separate, and a
single causal diamond will enclose both.  Thus, we are fundamentally
defining the range of a causal diamond in terms of the presence of
excitations in the matter sector.  This is well-defined in the entire
regime of semi-classical gravity, independently of the details of the
particle physics.

In a de~Sitter vacuum, the cosmological horizon has entropy given by
one quarter of its surface area.  The relevant area is the
cross-section of the top-cone of the causal diamond, which grows as
matter crosses the horizon. This production of Bekenstein-Hawking
entropy would contribute enormously to $\Delta S$.  However, it is
difficult to see the relevance of horizon entropy to the existence of
observers.  For the same reason, we will ignore the Bekenstein-Hawking
entropy produced in black hole formation.  This is also natural since
we have defined the causal diamond through this distinction.  As we
have emphasized above, horizons are a gravitational phenomenon and can
always be distinguished from the matter sector in the semiclassical
regime.  Hence, this restriction does not affect the universality of
our method.\footnote{A different question is whether the exclusion of
  black hole horizon entropy actually makes a qualitative difference
  to the results of this paper.  It seems likely that it would not.
  According to the Causal Entropic Principle, the preferred value of
  $\rho_\Lambda$ is set by the time of maximum entropy production.
  The growth of supermassive black holes, and of their entropy, is
  largest during the era a few Gyr after the big bang and eventually
  slows down.  Thus, it appears to set a similar timescale to the one
  we obtain from stellar entropy production.}

To summarize, we will consider exclusively the production of entropy
in ordinary matter.  We will weight a vacuum by the total entropy
increase in the causal diamond:
\begin{equation}
w=\Delta S~.
\label{eq-jkl}
\end{equation}

\paragraph{Motivation} 
From a pragmatic point of view, one can simply regard this proposal as
an attractive alternative to weighting by observers-per-baryon.  The
causal diamond is defined independently of low-energy physics, and the
entropy increase is a well-defined quantity that replaces more
specific assumptions about observers.  However, there are additional,
more fundamental reasons to embrace the Causal Entropic Principle.
Let us discuss each of the two conjectures put forward at the
beginning of this subsection.

The first conjecture is motivated by the study of black holes (see
\cite{Bou00a,Bou06,Bou06b} for details).  There is now considerable
evidence that black hole formation and evaporation is a unitary
process~\cite{StrVaf96,Mal97}.  This means that there will be two
copies of the initial state at the same instant of time, one inside
the black hole, and the other in the Hawking radiation outside.  This
would appear to violate the linearity of quantum mechanics.  However,
no actual observer can verify the problem~\cite{SusTho93,Pre92}, since
the two copies reside in causally disconnected regions of the
spacetime.  Hence, we can escape from the apparent paradox by
abandoning the global viewpoint, and be content with merely predicting
the observations of (any) one observer.

However, it would be unnatural for such a radical revision of our view
of spacetime to be confined to the context of black holes.  In many
cosmological solutions, a single observer can access only a small
portion of the global spacetime.  Hence, it is equally important to
restrict our attention to only one (cosmological) observer at a time,
which is what we do in this paper.  Descriptions of cosmology from the
local viewpoint can be found, for example, in
Refs.~\cite{Ban00,Bou00a,Bou00b,DysKle02,Bou05}.  Its implications for
eternal inflation were studied in Ref.~\cite{BouFre06}.

The second conjecture replaces far more specific conditions assumed to
be necessary for observers, such as the existence of galaxies, stable
planetary orbits, suitable chemistry, etc., which were needed in the
observers-per-baryon approach.  The basic idea is that observers,
whatever their form, have to obey the laws of thermodynamics.
Observation requires free energy and is clearly incompatible with
thermal equilibrium or an empty universe.  The free energy, divided by
the temperature at which it is radiated, can be regarded as a measure
of the potential complexity arising in a spacetime region.  This is
equal to the number of quanta produced, or more fundamentally, the
entropy increase $\Delta S$.

It seems plausible that there might be an absolute minimum complexity
necessary for observers, so that subcritical weights $\Delta S<\Delta
S_{\rm crit}$ should be replaced by zero.  For example, it seems
likely that vacua with $\rho_\Lambda$ of order unity, which can
contain only a few bits, have strictly zero probability of hosting
observers (see also Ref.~\cite{HarKri06}).  However, any such cutoff
does not appear to play an important role for the questions studied
here.  We choose the weight factor to be simply $\Delta S$, with no
cutoff.

To avoid confusion, we stress that our weighting has nothing to do
with the Hartle-Hawking amplitude, $\exp(S_{\rm dS})$, which describes
the number of quantum states associated with a de~Sitter
horizon~\cite{HarHaw83}.  The number of observers, and of
observations, is naturally proportional to an entropy difference,
$\Delta S$, not to an absolute entropy or the exponential of an
entropy.  (Despite our appropriation of the term ``entropic
principle'', for which we apologize to the authors, there is also
little relation to Ref.~\cite{hepth0509109}.)

We could have adopted only the first conjecture, and continued to
estimate the number of observers by more explicit anthropic criteria.
This would not have changed our final result significantly.  But why
make a strong assumption if a more conservative one suffices?  Our
results suggest that the poor predictions from weighting by
observers-per-baryon stemmed not from a lack of specificity in
characterizing observers, but from the regularization scheme (``per
baryon'').  The causal diamond cutoff solves this problem, and allows
us to weaken anthropic assumptions to the level of an inevitable
thermodynamic condition.

Moreover, $\Delta S$ is a well-defined weight in any vacuum, however
different from ours.  It will be a powerful tool in future work, when
parameters other than $\rho_\Lambda$ are allowed to vary
simultaneously.  We are thus motivated to use our second conjecture
throughout.  We will find that in our own universe, it captures
conditions for observers remarkably well.

\section{Computing $\rho_\Lambda$ from the Causal Entropic Principle}
\label{sec-calculation}

In this section, we will compute the probability distribution over
$\rho_\Lambda$, holding all other physical parameters fixed.  We do
this in four steps.  First, we will compute the geometry of the causal
diamond as a function of $\rho_\Lambda$.  Next, we will identify
important effects that produce entropy within the causal diamond.
Then we will determine the time-dependence of the entropy production
rate per unit comoving volume, as a function of $\rho_\Lambda$.
Finally, we will fold this together with the time-dependence of the
comoving volume contained in the causal diamond to obtain the weight
factor, $\Delta S(\rho_\Lambda)$.

\subsection{Metric and causal diamond}
\label{sec-metric}

Current data are consistent with a spatially flat universe.  Hence, we
will assume that since the time of reheating, the large scale
structure of our universe is described by a spatially flat FRW
model:
\begin{equation}
ds^2 = -dt^2 + a(t)^2 d {\mathbf x} ^2~.
\end{equation}
(Actually, we are making the stronger assumption that the cosmological
constant dominates before curvature does, for {\em all\/} values of
$\rho_\Lambda$ considered here.  Thus, we are assuming that the
universe is flatter than necessarily required by current constraints.)

After reheating, the universe will be dominated first by radiation,
then by matter, and finally by vacuum energy.  The reheating
temperature will not be important; in fact, we will neglect the
radiation era altogether.  Instead, we extrapolate the
matter-dominated era all the way back to the big bang ($t=0$), where
we will place the bottom tip of the causal diamond.  This
approximation is justified because the radiation era contributes only
a small fraction of conformal time in the range of values of $\rho_\Lambda$
that have any significant probability.  This is shown in detail in
Appendix~\ref{sec-radiation}.

Thus, we treat the universe as containing only pressureless matter and a
cosmological constant.  At early times (matter domination), the scale
factor will be proportional to $t^{2/3}$, independently of $\rho_\Lambda$.
At late times (vacuum domination) it will grow like
$\exp(t/t_\Lambda)$, where
\begin{equation}
t_\Lambda= \sqrt{\frac{3}{\Lambda}} =
\sqrt{\frac{3}{8\pi\rho_\Lambda}}~,
\label{eq-tlambda}
\end{equation}
(In our universe, with $\rho_\Lambda$ given by Eq.~(\ref{eq-cc}), we
have $t_\Lambda\approx 0.98 \times 10^{61} = 16.7$ Gyr).

An exact solution (aside from the neglected radiation era) that
includes both regimes is
\begin{eqnarray}
  a(t) & = & \left[t_\Lambda\sinh\left(\frac{3}{2}
      \frac{t}{t_\Lambda}\right)\right]^{2/3}\,,
\label{eq-aexact}\\
  \rho(t) & = & \rho_\Lambda+\rho_{\rm m} = 
  \rho_\Lambda \left[1 + \frac{1}{\sinh^2\left(\frac{3}{2}
        \frac{t}{t_\Lambda}\right)}\right]~.
\label{eq-rhoexact}
\end{eqnarray}
The prefactor, $t_\Lambda^{2/3}$, is arbitrary and can be changed by
rescaling the spatial coordinates.  Our choice ensures that for
solutions with different values of $\Lambda$, the scale factors will
agree at early times not only by diffeomorphism, but explicitly.  This
is convenient because $\Lambda$ is dynamically irrelevant at early
times.

Vacuum energy begins to dominate the density when $\rho_\Lambda
=\rho_{\rm m}$, at $t=0.59\, t_\Lambda$ (in our universe, 9.8 Gyr
after the big bang).  Acceleration begins earlier, when
$\rho+3p=\rho_{\rm m}-2\rho_\Lambda =0$, at $0.44\, t_\Lambda$ (in our
universe, at 7.3 Gyr).

In order to compute the boundaries of the causal diamond, it is
convenient to work in conformal time, $\tau=\int dt/a(t)$.  The metric
becomes
\begin{equation}
ds^2 = a(\tau)^2 [-d\tau^2+d {\mathbf x}^2]~.
\end{equation}
Light-rays obey $ds=0$, and hence for radial light-rays, $d\tau=\pm
dr$, where $r=|{\mathbf x}|$.

Using our solution, Eq.~(\ref{eq-aexact}), conformal time will be
given by
\begin{eqnarray}
\tau(t) & = & - t_\Lambda^{1/3}
\frac{1}{\cosh^{2/3}\left(\frac{3t}{2t_\Lambda}\right)}
F\left(\frac{5}{6},\frac{1}{3},\frac{4}{3},
\frac{1}{\cosh^2\left(\frac{3t}{2t_\Lambda}\right)}\right)\\ & = & -
\frac{t_\Lambda}{a(t)}
F\left(\frac{1}{3},\frac{1}{2},\frac{4}{3},
\frac{-t_\Lambda^2}{a(t)^3}\right)\,.
\end{eqnarray}
It has finite range, running from 
\begin{equation}
\tau(0) =
-\frac{\Gamma(\frac{4}{3})\Gamma(\frac{1}{6})}{\Gamma(\frac{1}{2})}
  t_\Lambda^{1/3} \approx -2.804\, t_\Lambda^{1/3}
\label{eq-tau0}
\end{equation}
at the big bang, to $\tau(\infty)=0$ at asymptotically late times.
The total conformal lifetime of the universe is $\Delta\tau=
\tau(\infty)-\tau(0) = 2.804\, t_\Lambda^{1/3}$.

The causal diamond is given by the region delimited by
\begin{equation}
r= \frac{\Delta\tau}{2}-\left|\frac{\Delta\tau}{2}+\tau\right|~.
\label{eq-cd}
\end{equation}
The comoving volume inside the causal diamond at conformal time $\tau$
is simply
\begin{equation}
\label{covol}
V_{\rm c}=\frac{4\pi}{3} r^3(\tau)\,.
\end{equation}
This is shown in Fig.~\ref{fig-covol}, as a function of proper time,
\EPSFIGURE[!t]{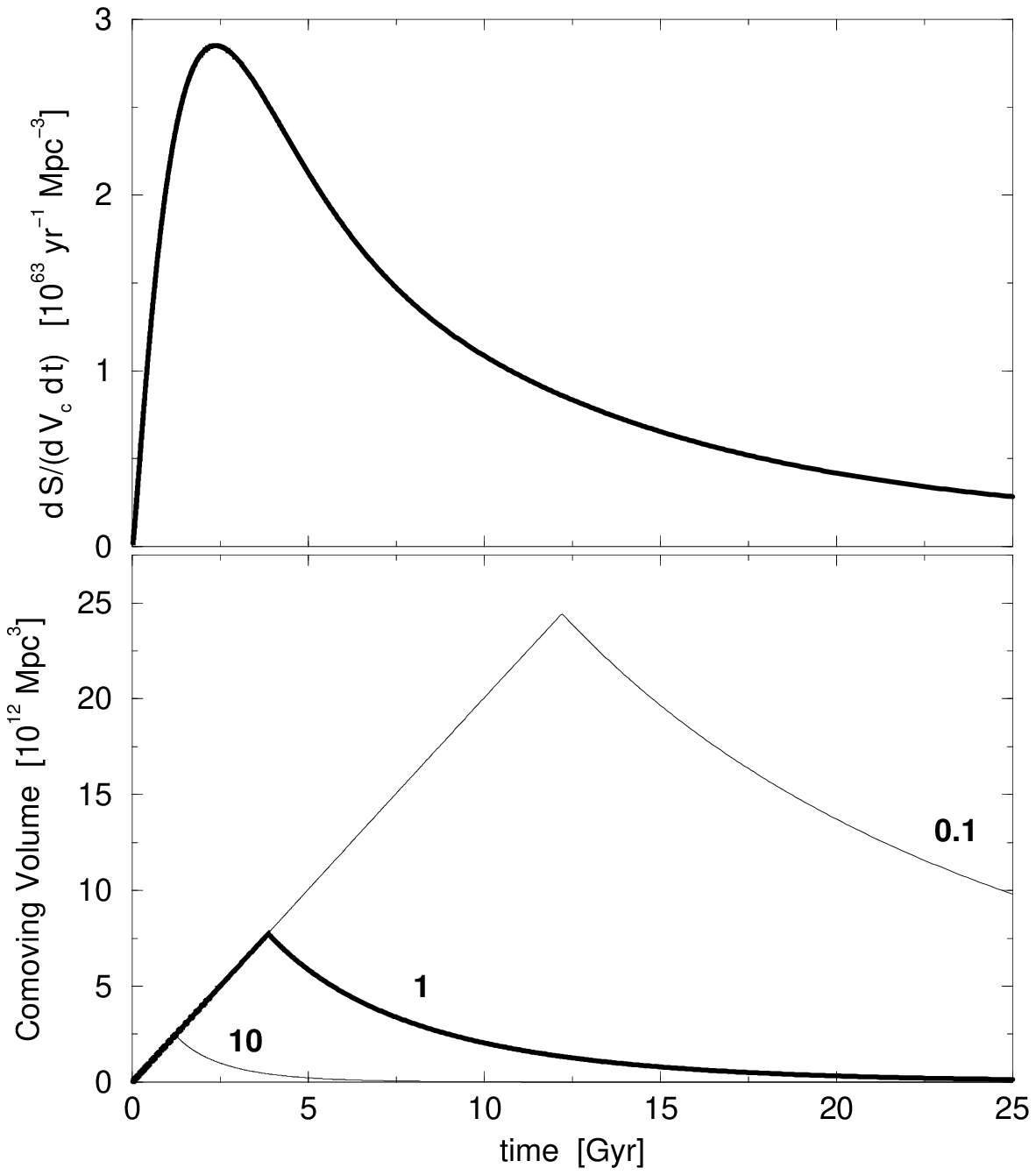,width=.8\textwidth}{\label{fig-covol} The
  lower plot shows comoving volume in the causal diamond, $V_{\rm c}$, as a function of time
  for $\rho_\Lambda = (0.1, 1, 10)$ times the observed value given in
  Eq.~(\ref{eq-cc}).  The kink in the comoving volume corresponds to
  the ``edge'' of the causal diamond, where the top and bottom cone meet (see Fig. 3).  The
  upper plot shows the rate of entropy production computed in
  Sec.~\ref{sec-rate}, which peaks around 2 to 3.5 Gyr.  As explained
  in Sec.~\ref{sec-understand}, the Causal Entropic Principle prefers
  values of $\rho_\Lambda$ such that the ($\rho_\Lambda$-dependent)
  peak of the comoving volume coincides with the (nearly
  $\rho_\Lambda$-independent) peak of the entropy production rate (see
  Fig.~9.}
for several values of $\rho_\Lambda$.  The maximum comoving volume 
occurs at the edge of the causal diamond at conformal time 
$\tau(0)/2$, or equivalently,
\begin{equation}
t_{\rm edge} \approx 0.23 t_{\Lambda}
\label{eq-edge}
\end{equation}
(in our universe, $t_{\rm edge} \approx 3.9$ Gyr).  The maximum
comoving volume itself is $V_{\rm c}(\tau(0)/2) \approx 11.6 t_\Lambda$.

At late times, the comoving volume goes to zero.  This reflects the
exponential dilution of all matter, which is expelled from the diamond
by the accelerated expansion.  The physical radius approaches a
constant, $t_\Lambda$, the horizon radius of the asymptotic de~Sitter
space.  Note that the ``comoving four-volume'',
\begin{equation}
V_4\equiv\int_0^\infty V_{\rm c}(t) dt~,
\end{equation}
is finite.  It is proportional to $t_\Lambda^2$, and hence, inversely
proportional to the cosmological constant.  This will be important,
since it means that smaller values of $\rho_\Lambda$ are rewarded by a
larger causal diamond, and thus, potentially greater complexity.  This
can compensate for their rarity.

\subsection{Major sources of entropy production}
\label{sec-entropy}

To calculate $dP/d(\log \rho_\Lambda)$, we need to calculate the total
entropy production in the causal diamond as a function of the
cosmological constant, $\Delta S(\rho_\Lambda)$.  We have determined
above how the causal diamond depends on $\rho_\Lambda$, but we must
also understand how the rate of entropy production depends on
$\rho_\Lambda$.  We begin by identifying the major sources
contributing to $\Delta S$ in the causal diamond in our own vacuum.

First, let us discuss sources which can be neglected.  Because the
causal diamond is small at early times ($\frac{\tau}{\tau_\infty}\ll
1$), and empty at late times ($1-\frac{\tau}{\tau_\infty}\ll 1$), the
most important contributions will be produced in the era between 0.1
Gyr to 100 Gyr, when the comoving volume shown in Fig.~\ref{fig-covol}
is large.  Hence, we can disregard the entropy produced at reheating,
at phase transitions, or by any other processes in the early universe.
Virtually all of this entropy (in particular, the cosmic microwave
background) entered the causal diamond through the bottom cone and
does not contribute to $\Delta S$.

For the same reason, we neglect the entropy in Hawking radiation
produced by supermassive black holes.  (One might contemplate the
possibility of a ``planet'' orbiting such an object and exploiting its
very low temperature radiation as a source of free energy.)  However,
the timescale for the evaporation of a black hole is enormous ($M^3$).
By the time this entropy would be produced, a typical causal diamond,
on which the measure for prior probabilities is based in the local
approach~\cite{Bou06}, will be completely empty.

Having dismissed effects at small comoving volume, we turn to
processes which operate from 0.1 Gyr to 100 Gyr, when the comoving
volume is large.  In this era, entropy is produced by baryonic
systems, and we can gauge the importance of a given process by its
total entropy production per baryon, or equivalently, per unit mass,
or unit comoving volume,\footnote{The choice of reference unit does
  not affect our weighting, because it drops out after integrating
  over the causal diamond.} $dS/dV_{\rm c}$. This can be estimated as
the ratio between the amount of energy released per baryon, and the
temperature at which this energy is dominantly radiated.

\paragraph{Stars} Ten percent of baryons end up in galaxies, and thus
in stars.  Approximately ten percent of those baryons actually burn in
the course of the lifetime of the star.  The energy released is about
7 MeV for each baryon in the reaction 4p $\rightarrow$ $^4$He, or about
$0.7 \times 10^{-2}$ of the rest mass of the proton.  Stars radiate at
varying temperatures corresponding largely to optical wavelengths
(about 0.2 to 3 eV).  However, more than half of this radiation is
reprocessed by dust~\cite{astroph9812182}.\footnote{We thank Eliot
  Quataert for pointing this out to us.}  It is re-emitted in the
infrared, at approximately $60\,\mu$m, or $20$ meV.  Hence, there will
be more than 100 infrared photons per optical photon~\cite{DolLag06},
and the infrared emission will dominate the entropy production.  In
summary, stellar burning produces an entropy of order $10^6$ per
baryon, after reprocessing by dust.

\paragraph {Active galactic nuclei} Active galactic nuclei (AGNs)
appear to be the main competitor to stars in terms of luminosity and
entropy production.\footnote{We thank Petr Horava for drawing our
  attention to AGNs.}  Approximately $10^{-3}$ of the total stellar
mass (i.e., $10^{-4}$ of baryonic mass) ends up in supermassive black
holes, and perhaps 10\% of this energy is released during the violent
accretion process.  This suggests that AGNs release at most about one
sixth of the energy as compared with stellar burning.
The effective temperature will again be 20 meV, since a large fraction
of the radiation is reprocessed by dust.  Hence, AGNs appear to
produce somewhat less total entropy per baryon than stars, and we will
neglect their contribution in the present paper.

In Ref.~\cite{Hopkins:2006fq}, the intrinsic luminosities of all AGNs
above observational limits were compiled to create a quasar luminosity
function applicable back to about 1 Gyr after the big bang.  This
luminosity function suggests that the entropy production rate from
quasars is more narrowly peaked than that estimated for stars in
Fig.~\ref{fig-dSdVdt}.  But even at the peak, near a redshift of
$z=2$, the quasar luminosity is a factor of 3 lower than the {\em
  present\/} stellar luminosity~\cite{Kashlinsky:2004jt} (which, in
turn, is about one order of magnitude lower than the peak stellar
luminosity).  This seems to rule out the possibility that AGNs ever
dominated the entropy production rate.

Other potentially important contributions come from galaxy cooling and
from supernovae.  Even assuming reprocessing by dust, neither of these
phenomena can compete with stellar burning, because they run on less
energy but not at lower temperature.

\paragraph{Supernovae} 
We focus on core collapse supernovae since they are more abundant than
type Ia supernovae (by factor of 4-5) while producing comparable
luminosity per event.  They occur in all stars with more than eight
solar masses, which constitute 1\% of the total stellar mass [see
  Eq.~(\ref{imf-eq})].  The collapse of an iron core into a neutron
star releases gravitational binding energy not much smaller than the
core mass (1.4 solar masses).  Thus roughly 10\% of the total mass of
the progenitor is released.  Most of this energy is carried away by
high energy neutrinos, producing little entropy.  Only 1\% produces
optical photons, and is reprocessed by dust into infrared photons.
Altogether, supernovae thus convert a fraction of $10^{-5}$ of stellar
baryonic mass into soft photons.

Further quantitative analysis confirms the above estimate.  We find
that entropy production from supernovae is more than order of
magnitude below the contribution from dust heated by stars.

\paragraph{Galaxy cooling} A typical galaxy like ours has a
mass-to-radius fraction of approximately $10^{-6}$.  This fraction of
the galaxy mass is converted into kinetic energy at virialization.
This energy, about $1$~keV per stellar baryon, is converted into
radiation as the galaxy cools.  The virial temperature is about $10^5$
K, or 10 to 100 times greater than the temperature of a star.  Even
assuming reprocessing by dust, galaxy cooling will produce less than
$10^4$ photons per baryon.  This is more than two orders of magnitude
below the entropy production from dust heated by stars.

\subsection{Entropy production rate}
\label{sec-rate}

We have argued that the dominant source of entropy production in our
universe is dust heated by starlight.  In this subsection we will
consider the rate at which this entropy is produced.  We will ask how
it depends on time and on $\rho_\Lambda$.

\paragraph{Time dependence} 
Deriving $dS/dV_{\rm c}\, dt$ from first principles would require a
detailed description of all of the dynamics that led up to stellar
burning, such as non-linear evolution, gas cooling and disk
fragmentation.  Instead, we will take advantage of observations that
quantify the time-dependence of the star formation rate.  This will
allow us to obtain the entropy production from dust heated by stars.
We will then estimate how this rate changes in universes with
different cosmological constants.  Surprisingly, this latter
dependence will not be important for our final result.

In recent years, observations of the extragalactic background
radiation in a large range of wavelengths have improved our
understanding of the galaxy luminosity function.  This has allowed
astronomers to infer the star formation rate (SFR) in galaxies, and
its evolution in time~\cite{Madau:1996yh,Lilly:1996ui}
(see~\cite{Kennicutt:1998zb} for a review).  The SFR is defined as the
rate of stellar mass production per comoving volume
\begin{equation}
  \dot \rho_\star (t)\equiv 
  \frac{\partial^2 M_\star}{\partial V_{\rm c} \partial t}\,.
\end{equation}
This function is constrained by observation through a variety of
techniques. For example, UV emission from star-forming galaxies is
dominated by massive stars that are short-lived. Due to the short
lifetime, luminosities in these wavelengths track the birth rate of
stars~\cite{Madau2}.  In addition, detailed surveys of the local
universe constrain the SFR at low
redshift~\cite{Cole:2000ea,Hopkins:2003am}.  Bounds on the rate of
type II supernovae from Super-Kamiokande and KamLAND also place an
indirect bound on stellar birth~\cite{Fukugita:2002qw}.

The combination of these measurements constrain the SFR back to
redshifts of about $z\sim 6$, when the universe was 1 Gyr old. Since
then, the SFR may be grossly described as a smooth function that peaks
at around 2.5 Gyr and subsequently decreases exponentially with
time. The SFR today is roughly one order of magnitude less than its
peak value. As we shall see, the era of peak stellar formation, which
we will call $t_\star$, will play an essential role in determining the
cosmological constant using the Causal Entropic Principle.

Several authors have postulated models or functional forms for the
SFR, fitting model parameters to the data, for
example~\cite{Cole,Gal-Yam:2003ed,Nagamine:2006kz,Hopkins:2006bw}.
Variations between the fits lead to a range for $t_\star \sim 1.5
- 3$ Gyr.  

We will illustrate our computation using two different SFRs, in order
to illustrate the systematic dependence of our calculation on the time
dependence of star formation.  The first SFR is from
Ref.~\cite{Nagamine:2006kz} (labeled ``N'' in plots) and has
$t_\star\sim 1.7$ Gyr. The second SFR, from Ref.~\cite{Hopkins:2006bw}
(labeled ``H''), peaks at $t_\star\sim 2.8$ Gyr.  We will find in both
cases that the observed value of $\rho_\Lambda$ lies well inside the
$1\sigma$ region of the resulting probability distribution.

\EPSFIGURE[!t]{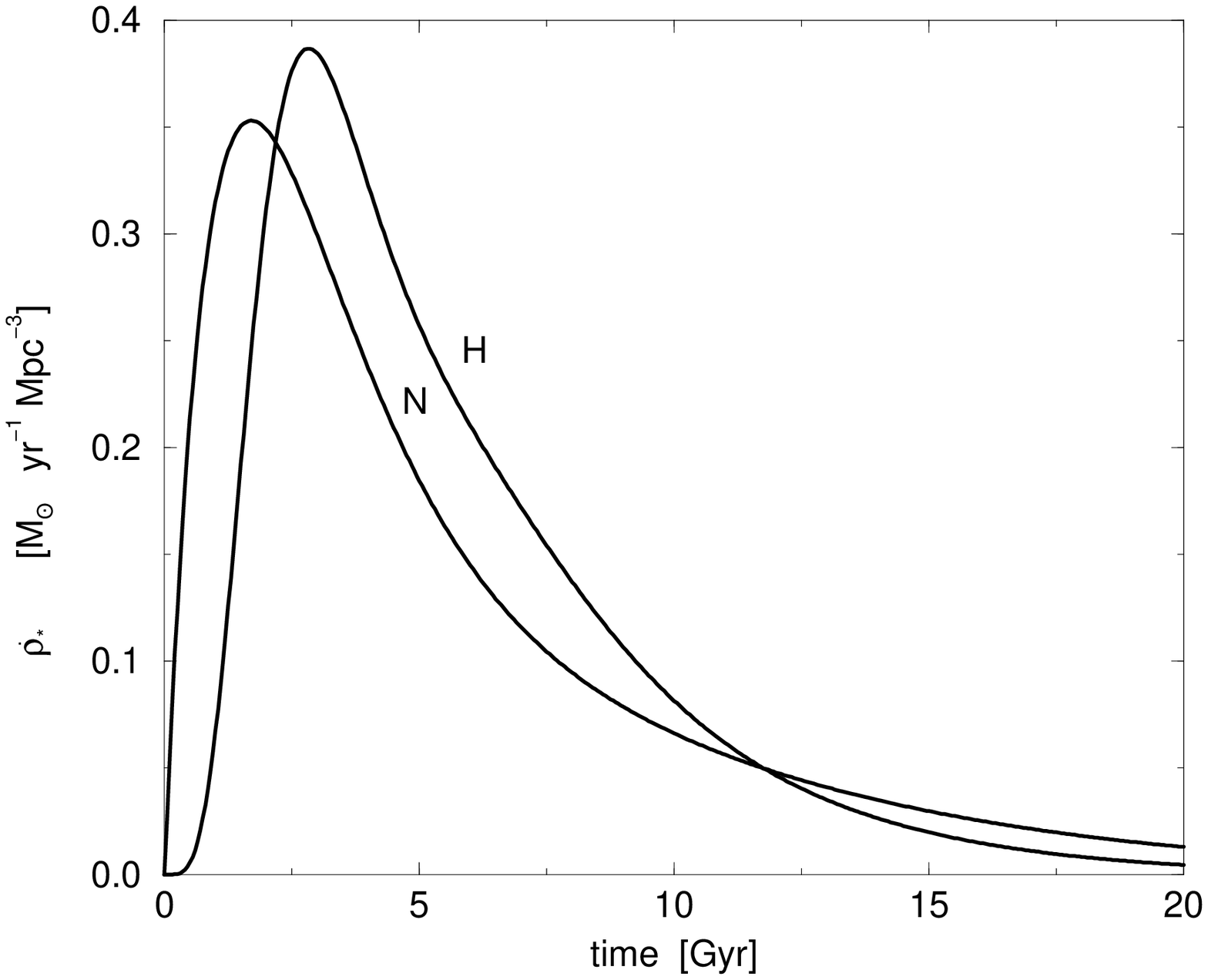,width=.8\textwidth}{\label{fig-SFR} 
The star
  formation rate as a function of time from Nagamine {\em et
    al.}~\cite{Nagamine:2006kz}, labeled ``N'', and from Hopkins and
Beacom~\cite{Hopkins:2006bw}, labeled ``H''.
They are peaked at $t_\star\sim$1.7 Gyr and 2.8
  Gyr respectively. We have normalized both SFRs such that the stellar
  luminosity density calculated below agrees with the bolometric
  luminosity observed today. It should be noted, however, that our
  result is not sensitive to this re-normalization.}
Both SFRs are shown in Fig.~\ref{fig-SFR} re-normalized.  One of the
biggest uncertainties regarding the determination of the star
formation rate is the overall normalization of the curve. Fortunately
our result is \emph{completely} insensitive to this overall
normalization; it depends only on the shape of the SFR.

For concreteness, we have normalized the SFR such that the implied
stellar luminosity today matches the observed bolometric luminosity
from low-redshift stars, $10^{8.6} L_\odot/\mathrm{Mpc}^3$.  Our
normalizations differs from those in the original works since we have
used rather rudimentary formulae to compute the present luminosity
from the SFR. However, the shape of the total entropy production rate
derived with these simple formulae agrees well with that in
Ref.~\cite{Nagamine:2006kz} when this SFR is used.

With the rate of star formation in hand, one can estimate the rate of
entropy production by stars. Let us first consider a single star of
mass $M$ born at a time $t'$. Its entropy production rate is the
stellar luminosity divided by the temperature at which photons are
radiated,
\begin{equation}
  \frac{d^2s}{dN_\star dt}(M)= \frac{L_\star}{T_{\rm eff}}\sim
  \frac{1}{2}\left(\frac{M}{M_\solar}\right)^{3.5}\frac{L_\solar}{20\,
    \mathrm{meV}} = \frac{1}{2}\left(\frac{M}{M_\solar}\right)^{3.5}
  3.7 \times 10^{54} \ \mathrm{yr}^{-1}\,.
\end{equation}
Here we have assumed a mass-luminosity relation 
\begin{equation}
\label{eq-masslum}
L_\star\propto M^{3.5}\,.
\end{equation}  
(We use $M_\star$ to refer to total stellar mass, and $M$ to refer to
the mass of a specific star.  $N_\star$ denotes star number.)  The
effective temperature of 20 meV is that of the dust which reprocesses
about one half (hence the prefactor) of the starlight and dominates
photon number.

The star will only produce entropy over a finite time $t_\star$ that
also depends on $M$:
\begin{equation}
\label{stellarlifetime}
t_\star (M)\sim \left(\frac{M_\solar}{M}\right)^{2.5} 10^{10} \; \mathrm{yr}.
\end{equation}   

Now let us consider a whole population of stars that are formed at a
time $t'$. At birth, stellar masses are observed to be distributed
according to a universal, time-independent function known as the
initial mass function (IMF).  The Salpeter
distribution~\cite{Salpeter:1955it}, which goes as $M^{-2.35}$, agrees
reasonably with observation, but has since been refined by many
authors.  In the present calculation we will use a modified Salpeter
IMF of the form~\cite{Hopkins:2006bw}
\begin{equation}
\xi_{\mathrm{IMF}}(M)\equiv \frac{dN_\star}{dM} = \left\{ 
\begin{array}{ll}
C_1M^{-2.35} & \mbox{for}\ M\ge 0.5 M_\solar \\ 
C_2M^{-1.5} & \mbox{for}\ M<0.5 M_\solar
\end{array} \right.~, \label{imf-eq}
\end{equation}
where the constants $C_{1,2}$ are set by requiring that the IMF is
continuous and that it integrate to one over the range $0.08
M_\solar<M<100 M_\solar$.

We now have all the ingredients in place to calculate $d^2S/(dM_\star
dt)$, the contribution of a stellar population that is born at time
$t'$ to the entropy production rate at some later time $t>t'$, per
unit initial stellar mass. This rate is a function only of the time
difference $t-t'$
\begin{equation}
  \frac{d^2S}
  {dM_\star dt} (t-t') =\frac{1}{\langle M \rangle}
  \int_{M_{\mathrm{min}}}^{M_{\mathrm{max}}(t-t')} dM\,
    \xi_{\mathrm{IMF}} \frac{d^2s}{dN_\star
      dt}(M)~,
\label{eq-lkj}
\end{equation}
with the average initial mass (defined at $t'$), $\langle M \rangle$,
defined as
\begin{equation}      
\langle M \rangle = \int_{0.08 M_\odot}^{100 M_\odot} dM\,
\xi_{\mathrm{IMF}} M\approx 0.48 M_\odot ~.
\end{equation}
The time dependence enters Eq.~(\ref{eq-lkj}) through the death of stars
of various masses at different times and thus appears in the upper
limit of the upper integral.  It is derived by inverting
equation~(\ref{stellarlifetime}):
\begin{equation}
  M_{\mathrm{max}}(t-t')=\left\{ 
  \begin{array}{ll}
   100 M_\solar & \ \mbox{for}\ t-t'<10^5\ \mbox{yr} \\   
   \left(\frac{10^{10} \; \mathrm{yr}}{t-t'}\right)^{2/5} M_\solar&
      \ \mbox{for}\ t-t'>10^5\ \mbox{yr} \\
  \end{array}\right.\,.
\end{equation}
The lower limit of the integral is set by the minimal mass of stars
that can sustain nuclear burning,
\begin{equation}
M_{\mathrm{min}}=0.08 M_\solar\,.
\end{equation}
These stars burn for $\sim 5000$ Gyr, living well into vacuum domination in
our universe.

\EPSFIGURE[!t]{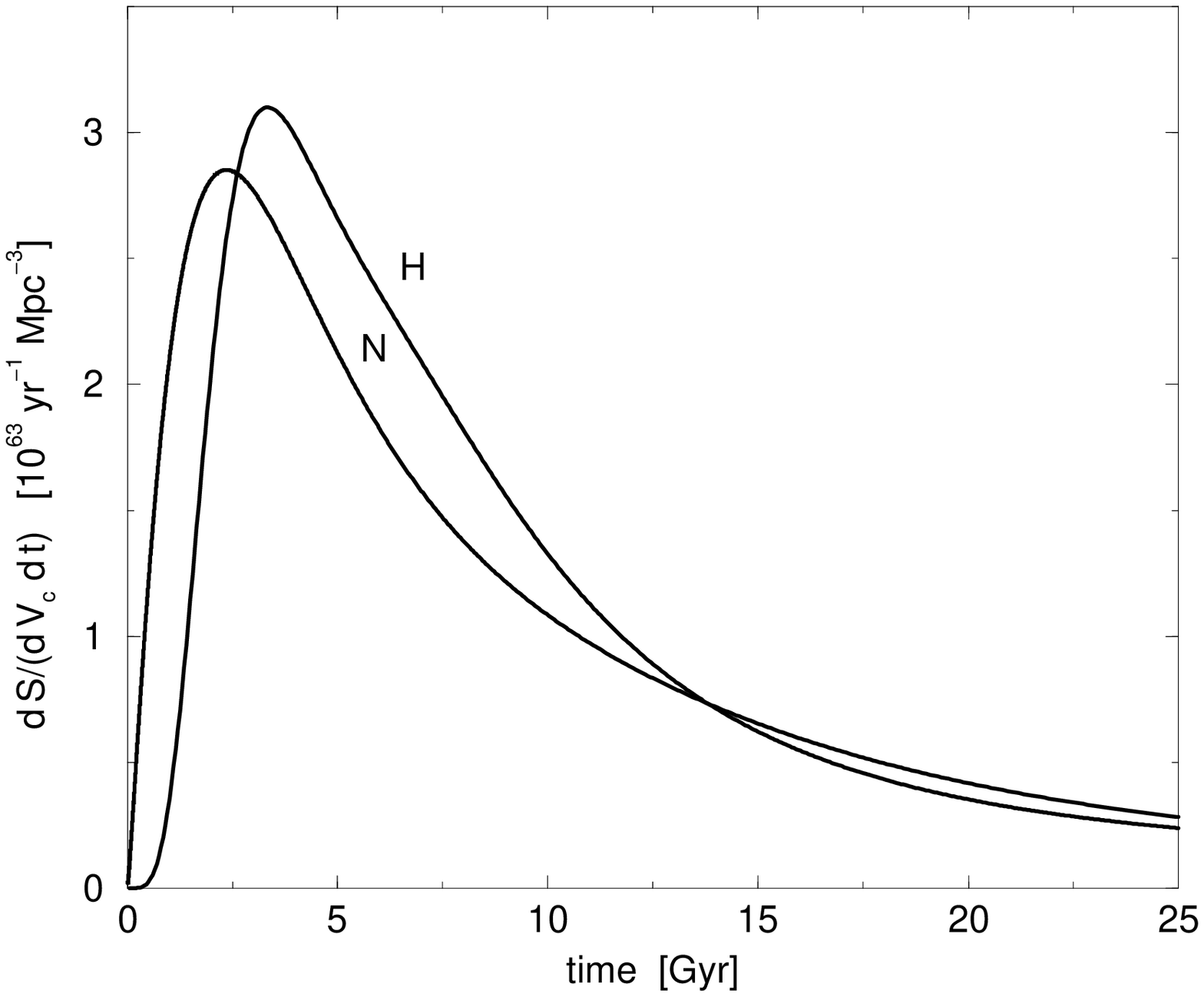,width=.8\textwidth}{\label{fig-dSdVdt} The
  entropy production rate in our universe as a function of time, from
  dust heated by starlight.  The two curves shown correspond to
  different models of the star formation history of the
  universe~\cite{Nagamine:2006kz,Hopkins:2006bw}; see Fig.~5.}
In order to calculate the entropy production rate in the universe at
time $t$ we must convolve $d^2S/(dM_\star dt)$ with the SFR.  That is,
we integrate over the birth times $t'$ of all populations of stars
born before the time $t$:
\begin{eqnarray}
\frac{d^2 S}{d V_{\rm c} d t}(t)&=&
\int_0^t dt'\, \frac{d^2 S}{d M_\star d t}(t-t') 
\frac{d^2 M_\star}{d V_{\rm c} d t'}(t') \nonumber \\
&=& \int_0^t dt'\,\frac{d^2 S}{d M_\star d t}(t-t')\, \dot \rho_\star(t')\,.
\label{dSdVdt}
\end{eqnarray}

This function is plotted in Fig.~\ref{fig-dSdVdt} for the two forms of
the SFR.  It is a smooth function that peaks when the universe is
about 2.3 Gyr old and 3.3 Gyr old for the two curves plotted. This
timescale is set by the peak of the star formation rate and the mean
lifetime of a star in our universe.  The entropy production rate
decreases as stars born during the peak of the SFR begin to die.  But
due to the high abundance of long-lived low mass stars,
$\partial_{V_{\rm c}} \dot S$ maintains a finite value long after star
formation has ceased.

In our approximation, the entropy production rate is half the
luminosity of stars at the time $t$, divided by the effective
temperature (dust at 20 meV).  Modulo this rescaling,
Fig.~\ref{fig-dSdVdt} thus also shows our estimate for the luminosity,
which was used for normalization as described above.

We caution that we assumed overly simplistic formulae for the
luminosity and the lifetime of a star.  For example, we ignored the
dependence on metallicity, and a dependence of the exponent in
Eq.~(\ref{eq-masslum}) on the mass.  This will likely affect the shape
of the entropy production rate somewhat, as will corrections from
other sources of entropy.  However, our prediction for $\rho_\Lambda$
depends only on the roughest features: the width and position of the
peak of the entropy production rate.  Hence, it is unlikely that our
result would be qualitatively affected by our simplifications.

\paragraph{Dependence on $\rho_\Lambda$}
The entropy production rate calculated above is that in our universe.
In order to determine $\Delta S(\rho_\Lambda)$, we will need to
calculate this function for universes with different values of the
cosmological constant.  Interestingly, this dependence will be of
little importance for our final result.

Stars have decoupled from the global expansion of the universe, so
their internal dynamics is unaffected by variations of $\rho_\Lambda$.
However, the value of $\rho_\Lambda$ affects the fraction of baryons
that are incorporated in halos large enough to form stars at any given
time, thus affecting the star formation rate, and ultimately the
entropy production.  For example, if $\rho_\Lambda$ is large enough to
violate Weinberg's anthropic bound, no baryons will be in star forming
halos and $\Delta S$ will be very suppressed.

In a universe with a vacuum energy density $\rho_\Lambda$, the
fraction of matter that is incorporated in halos of a mass $M_G$ by
time $t$ or above, $F(\rho_\Lambda, M_G, t)$, is easily calculated in
the Press-Schechter (PS) formalism~\cite{Press:1973iz}. The formulae
for the PS fraction are summarized in~\cite{Tegmark:2005dy}.  (This
fraction is also a function of the amplitude of density perturbations
and the temperature at matter-radiation equality, but since these are
held fixed in this calculation we will suppress them.)  

For the purpose of our calculation, we will consider a galaxy to be
star-producing if the mass of its host halo is $10^7 M_\solar$ or
above.  Note that this choice involves no speculation about what
observers need.  The Causal Entropic Principle requires us to compute
the entropy production rate as a function of $\rho_\Lambda$.  This
rate depends on whether or not stars, the dominant contributors of
entropy, actually form.  For $M_G>10^7 M_\odot$, the virial
temperature of the halo is above $10^3$ K, enough to support rapid
line cooling and efficient stellar production.  (The first generation
of stars---Population III---were formed in galaxies with even lower
masses, but since these stars have been neither observed nor accounted
for in the observation-based SFRs, we will not consider them further.)
In any case, as we shall see, our final result will be quite
insensitive to this choice.

Based on the SFR $\dot \rho_\star (t)$ in our universe, we will now
estimate the SFR in a universe with a different vacuum energy, $\dot
\rho_\star (\rho_\Lambda,t)$.  The SFR in our universe is peaked at
about $t_\star\sim 1.5-3$ Gyr, which is still in the matter dominated
era. The cosmological constant played no dynamical role and cannot
have anything to do with this peak.  Rather, this time scale is set by
astrophysical dynamics, such as galaxy formation, cooling, and
feedback.  

\EPSFIGURE[!t]{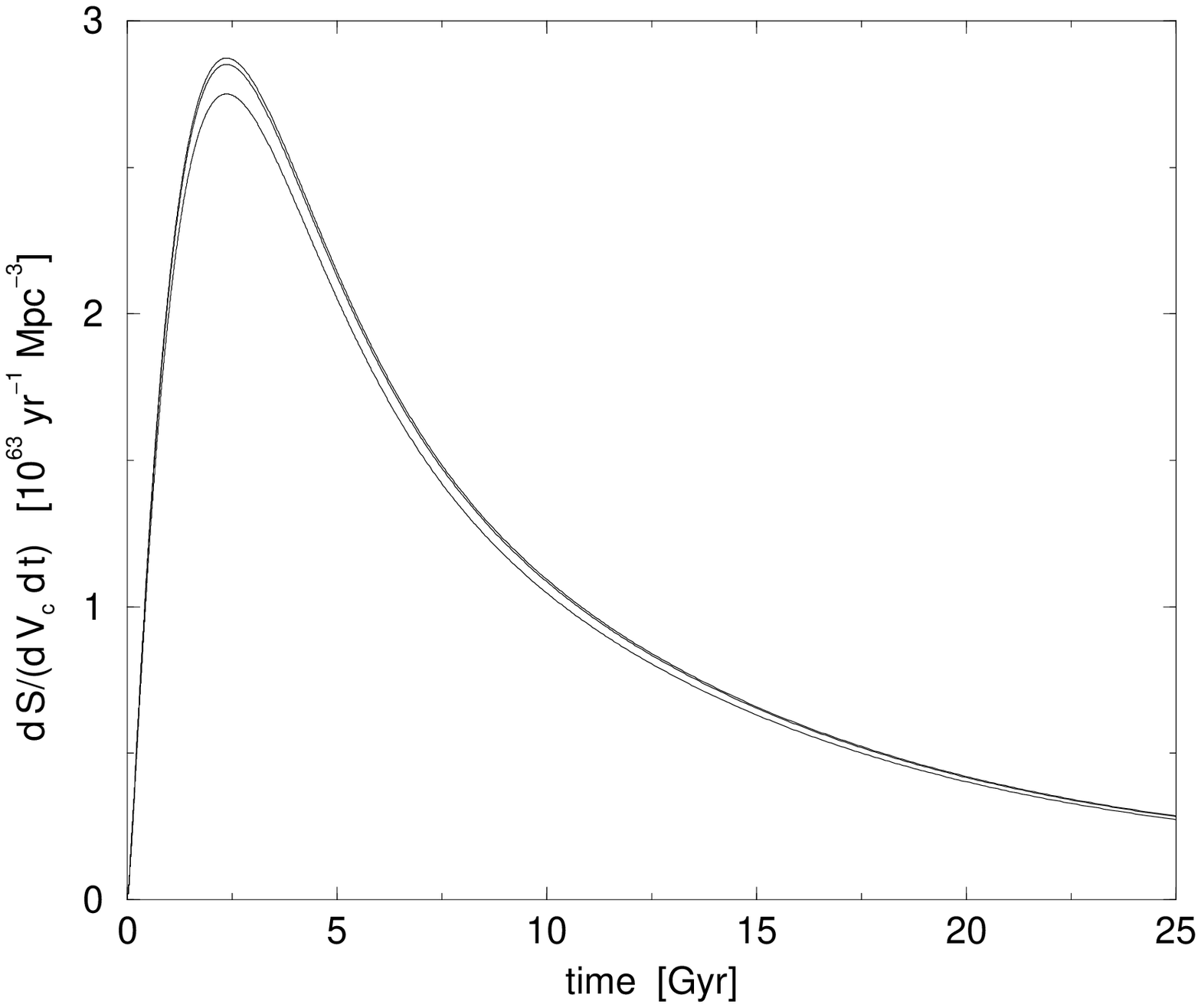,width=.8\textwidth}{\label{fig-eprrl} The
  entropy production rate of Fig.~\ref{fig-dSdVdt} depends only mildly
  on the vacuum energy.  Hence, dependence on the vacuum energy enters
  our calculation mainly through the size of the causal diamond (see
  Figures 4 and 9).  The rate of Ref.~\cite{Nagamine:2006kz} is shown
  here for $\rho_\Lambda = (0.1, 1, 10)$ times the observed value
  (from top to bottom) using the approximation of
  equation~\ref{fig-approx}.}
Therefore, the star formation rate depends mainly on the mass fraction
in star forming galaxies at the critical time $t_\star$.  Strictly
speaking, the SFR will be sensitive to the PS fraction at times before
$t_\star$ because of the cooling period that is required between the
time a baryonic structure goes non-linear and the time it collapses
into stars.  Baryons that burn during the peak of the SFR actually
fell into non-linear halos a cooling time earlier.  We leave a more
careful analysis of these and other effects to further investigation.

Let us define $F_\star (\rho_\Lambda)$ as the PS fraction that is most
relevant for star formation in a universe with cosmological constant
$\rho_\Lambda$, namely that evaluated at $t_\star$, with a minimum
mass of $10^7 M_\solar$.  In order to capture the mild sensitivity of
the SFR to changes of the cosmological constant, we rescale the SFR by
the appropriate $F_\star$:
\begin{equation}
\label{fig-approx}
\dot \rho_\star (\rho_\Lambda, t)= 
\dot\rho_\star (t)\times
\frac{F_\star(\rho_\Lambda)}
{F_\star(1.25\times 10^{-123})}\,.
\label{eq-sfrps}
\end{equation}

Because the observed value of $\rho_\Lambda$ is far from disturbing
the formation of $10^7 M_\odot$ galaxies, small variations of
$\rho_\Lambda$ will barely affect the mass fraction $F_\star$; see
Fig.~\ref{fig-eprrl}.  But they do affect the geometry of the causal
diamond.  This is the reason why the latter effect will be important
for our final result, while Eq.~(\ref{eq-sfrps}) gives only a tiny
correction.

\subsection{Total entropy production in the causal diamond}
\label{sec-total}

From the above results, we can compute the total entropy production
\begin{equation}
  \Delta S(\rho_\Lambda) = 
  \int_0^\infty dt\, V_{\rm c}(\rho_\Lambda,t) \partial_{V_{\rm c}} \dot S
  (\rho_\Lambda,t)~.
\end{equation} 
Here, $V_{\rm c}$ is the comoving volume in the causal diamond at the time
$t$, given in Eq.~(\ref{covol}).  $\partial_{V_{\rm c}} \dot S$ is the rate
of entropy production per unit comoving volume, given in
Eq.~(\ref{dSdVdt}).  The dependence on $\rho_\Lambda$ enters mainly
through $V_{\rm c}$.  It is straightforward to perform the integrals
numerically.

By Eqs.~(\ref{eq-plog}) and (\ref{eq-jkl}), the probability
distribution $dP/d(\log \rho_\Lambda)$ is proportional to
$\rho_\Lambda \Delta S(\rho_\Lambda)$.  We show this distribution in
Fig.~\ref{fig-ae} for both the Nagamine {\em et
  al.}~\cite{Nagamine:2006kz} as well as the Hopkins {\em et
  al.}~\cite{Hopkins:2006bw} SFRs.
\EPSFIGURE[!t]{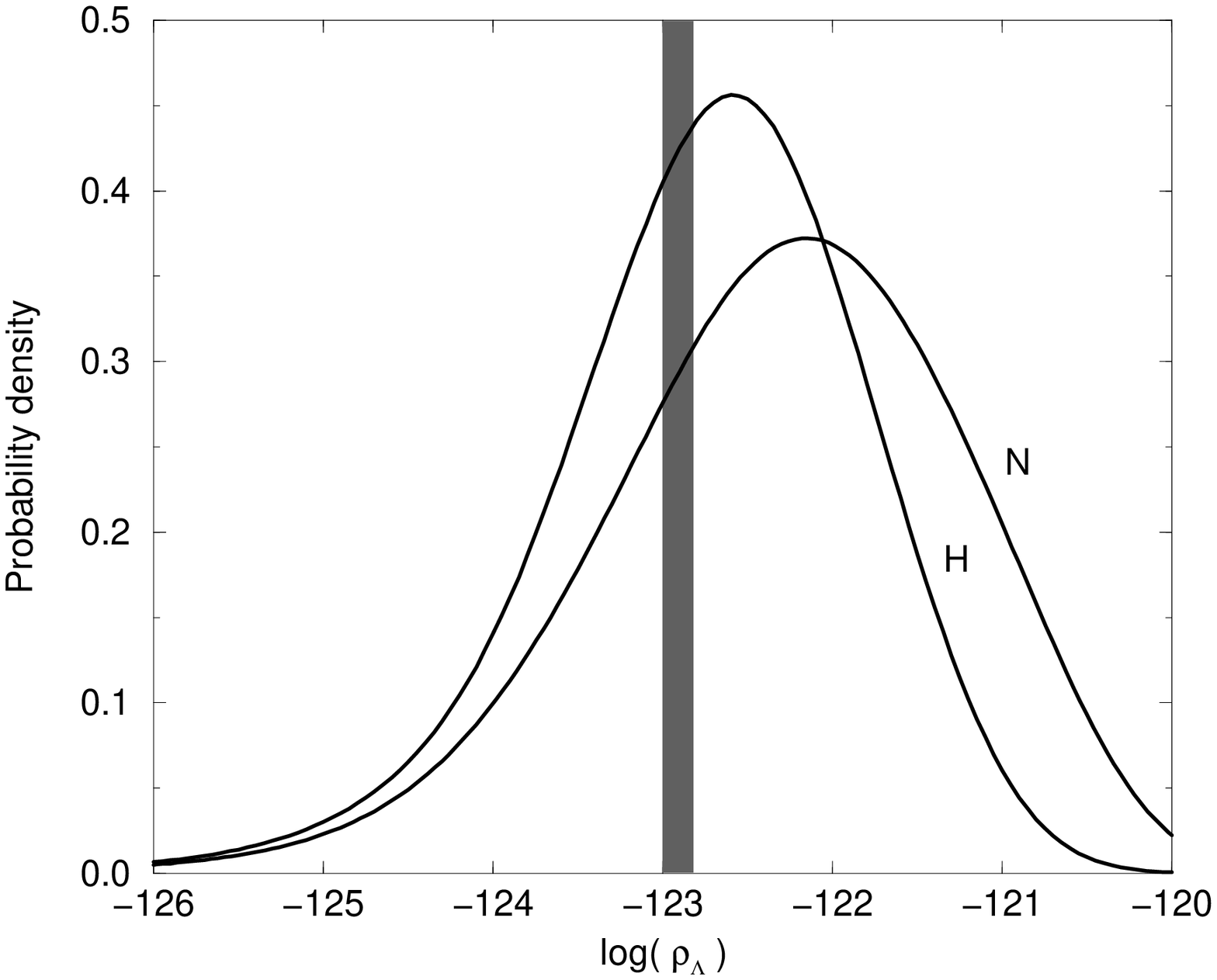,width=.8\textwidth}{\label{fig-ae}
  The probability distribution over $\log(\rho_\Lambda)$ computed from
  the Causal Entropic Principle.  The two curves shown arise
  from two different models of the star formation rate (see
  Figs.~\ref{fig-SFR} and~\ref{fig-dSdVdt}).  Their differences hint
  at the systematic uncertainties in our calculation that arise since
  the history of entropy production is not known to arbitrary
  precision.  These uncertainties are apparently irrelevant to our main conclusion:
  either way, the observed value of $\rho_\Lambda$ (vertical bar) is
  not unlikely.}
For the Nagamine {\em et al.} SFR, the median value is $\rho_\Lambda
= 5.6 \times 10^{-123}$ with the $1\sigma$ error band between $4.2
\times 10^{-124}$ to $5.8 \times 10^{-122}$.  For the Hopkins et
al.\ SFR, we find the median value $\rho_\Lambda = 2.1 \times
10^{-123}$ with the $1\sigma$ error band between $2.4 \times
10^{-124}$ to $1.4 \times 10^{-122}$.

There are ``systematic uncertainties'' in our calculation of the
probability distribution, which come from our lack of knowledge of the
precise history of entropy production in our universe.  The comparison
between the two models for the SFR gives a good sense of their size.
It shows that these uncertainties do not affect our conclusion: with
either choice, the observed value is well within the $1\sigma$ region,
and hence, not unlikely.

\section{Discussion}
\label{sec-discussion}

\subsection{$\Delta S$ captures complexity}
\label{sec-deltas}

Our main quantitative result is the probability distribution for
$\rho_\Lambda$.  However, we have also discovered an important
intermediate result: in our own causal diamond, the dominant
contribution to entropy production comes from the infrared glow of
dust heated by starlight.\footnote{It is amusing to note that this
  class of entropy producers includes the authors and the reader, in
  the sense that the Earth, like dust, is made of heavy elements and also 
  absorbs starlight and re-emits in the infrared.}

This is remarkable.  It shows that a seemingly primitive quantity,
$\Delta S$, captures many of the conditions that are usually demanded
explicitly by anthropic arguments.  $\Delta S$ would drop sharply if
galaxies, stars, or heavy elements were absent.  According to the
Causal Entropic Principle, the weight carried by such a vacuum would
be suppressed.

For example, consider a universe like ours, except without heavy
elements.  (This could be arranged by adjustments in the standard
model.)  Galaxies would still form, and stars would burn, but there
would be no dust available to convert optical photons into a much
larger number of infrared photons~\cite{astroph9812182}.  The Causal
Entropic Principle assigns a weight 100 times larger to our vacuum
than to this one---simply based on the entropy production, without
knowing anything about the potential advantages of heavy elements
often claimed in anthropic arguments.

This demonstrates that $\Delta S$ can be an effective and
very simple substitute for a number of dubious anthropic conditions.
More importantly, our result lends credibility to $\Delta S$ as a
weighting factor for vacua with very different low-energy physics.
Estimating the number of observers in such vacua, even averaged over a
large parameter space, appears wholly intractable, but estimating
$\Delta S$ may not be.  

Thus, the Causal Entropic Principle may allow us, for the first time,
to predict probability distributions over the entire landscape, rather
than just over subspaces constrained to coincide with much of our
low-energy physics.  As discussed in Sec.~\ref{sec-intro}, this could
lead to a breakthrough on extracting predictions directly from the
underlying theory (the string landscape), without conditioning on
parameters specific to our own vacuum.

\subsection{Understanding our distribution}
\label{sec-understand}

Now, let us turn to our main result.  In our approach, the most likely
range of $\log\rho_\Lambda$ is set not by the time of galaxy
formation, but by the time at which the {\em rate\/} of entropy
production, per unit time and unit comoving volume, is largest.  This
can be understood as follows.

Consider, for the sake of argument, an entropy production rate that is
independent of time and of $\rho_\Lambda$.  Then the total entropy
$\Delta S$ produced within the causal diamond is proportional to $\int
V_{\rm c}(t)\, dt$, where $V_{\rm c}$ is the comoving volume (or
equivalently, the mass) present inside the causal diamond at the time
$t$.  This integral is the area under the curves shown in the lower
panel of Fig.~\ref{fig-covol}.

At small times (near the bottom tip) the causal diamond is small and
$V_{\rm c}$ is negligible.  After vacuum domination, at a time of order
$t_\Lambda=(3/8\pi\rho_\Lambda)^{1/2}$, the top cone, which contains
one de~Sitter horizon region, quickly empties out and $M(t)$ vanishes
exponentially.  Thus, only the era around the time of matter/vacuum
equality contributes significantly to the integral.

Up to a $\rho_\Lambda$-independent factor of order unity, the above
integral is therefore the product of $t_\Lambda M(t_\Lambda)$.  But
$M(t_\Lambda)$ is just the total mass inside the horizon around the
time when $\Lambda$ begins to dominate.  This is again of order
$t_\Lambda$ and thus proportional to $\rho_\Lambda^{-1/2}$.  Hence the
total entropy produced, $\Delta S$, scales like $1/\rho_\Lambda$ in
our hypothetical case of a constant entropy production rate.  This is
also clear by inspecting the area under the different curves in
Fig.~\ref{fig-covol}.

Assuming a flat prior ($dp/d \rho_\Lambda = \mbox{const}$, or
equivalently, $dp/d\log \rho_\Lambda \propto \rho_\Lambda$), the
observer-weighted probability distribution is
\begin{equation}
\frac{dP}{d\log \rho_\Lambda} \propto w(\rho_\Lambda ) \rho_\Lambda ~,
\end{equation}
and the Causal Entropic Principle states that the weight is
\begin{equation}
w=\Delta S~.
\end{equation}

For the hypothetical, constant entropy production rate, we have
$w\propto \rho_\Lambda^{-1}$, and hence
\begin{equation}
\frac{dP}{d\log \rho_\Lambda}  = \mbox{const} ~.
\end{equation}
The weight $\Delta S$ in this case takes a prior distribution that was
flat in $\rho_\Lambda $ into an observer-weighted distribution that is
flat in $\log \rho_\Lambda $, showing no preference between, say,
$\rho_\Lambda =10^{-121}$ and $\rho_\Lambda = 10^{-123}$.

In the prior distribution, there are more vacua at large
$\rho_\Lambda$, so exponentially small values of $\rho_\Lambda $ are
very unlikely.  The above example shows that the Causal Entropic
Principle captures an important compensating factor: vacua with
smaller $\rho_\Lambda$ give rise to a larger causal diamond, i.e., to
a bigger de~Sitter horizon and a longer time until vacuum domination.
This allows for greater complexity and compensates for the rarity of
such vacua.

Next, let us consider the time-dependent entropy production rate we
found in Sec.~\ref{sec-rate}.  We found that the entropy production
due to stars has a fairly broad peak around $t_{\rm peak}\sim 2$ to 3.5 Gyr
after the big bang.  At earlier times, it is lower because fewer stars
have formed; at late times, it is lower because few new stars form
while older ones have burned out.

Because of the time dependence, $\rho_\Lambda \Delta S$ will no longer
be constant.  For large values of $\rho_\Lambda$, the causal diamond
is small, and it will contain only a small comoving volume by the time
the entropy production peaks [$t_{\rm peak}\gg t_{\rm edge}=0.23
t_\Lambda$; see Eq.~(\ref{eq-edge})].  In this regime, $dP/d\log
\rho_\Lambda$ will decrease with $\rho_\Lambda $.  For small
$\rho_\Lambda$, the causal diamond is very large, but the entropy
production rate peaks early, when the comoving volume is still
relatively small ($t_{\rm peak}\ll t_{\rm edge}$).  In this regime,
$dP/d\log\rho_\Lambda$ will increase with $\rho_\Lambda $.  This is
illustrated in Fig.~\ref{fig-goldilocks}.

\EPSFIGURE[!t]{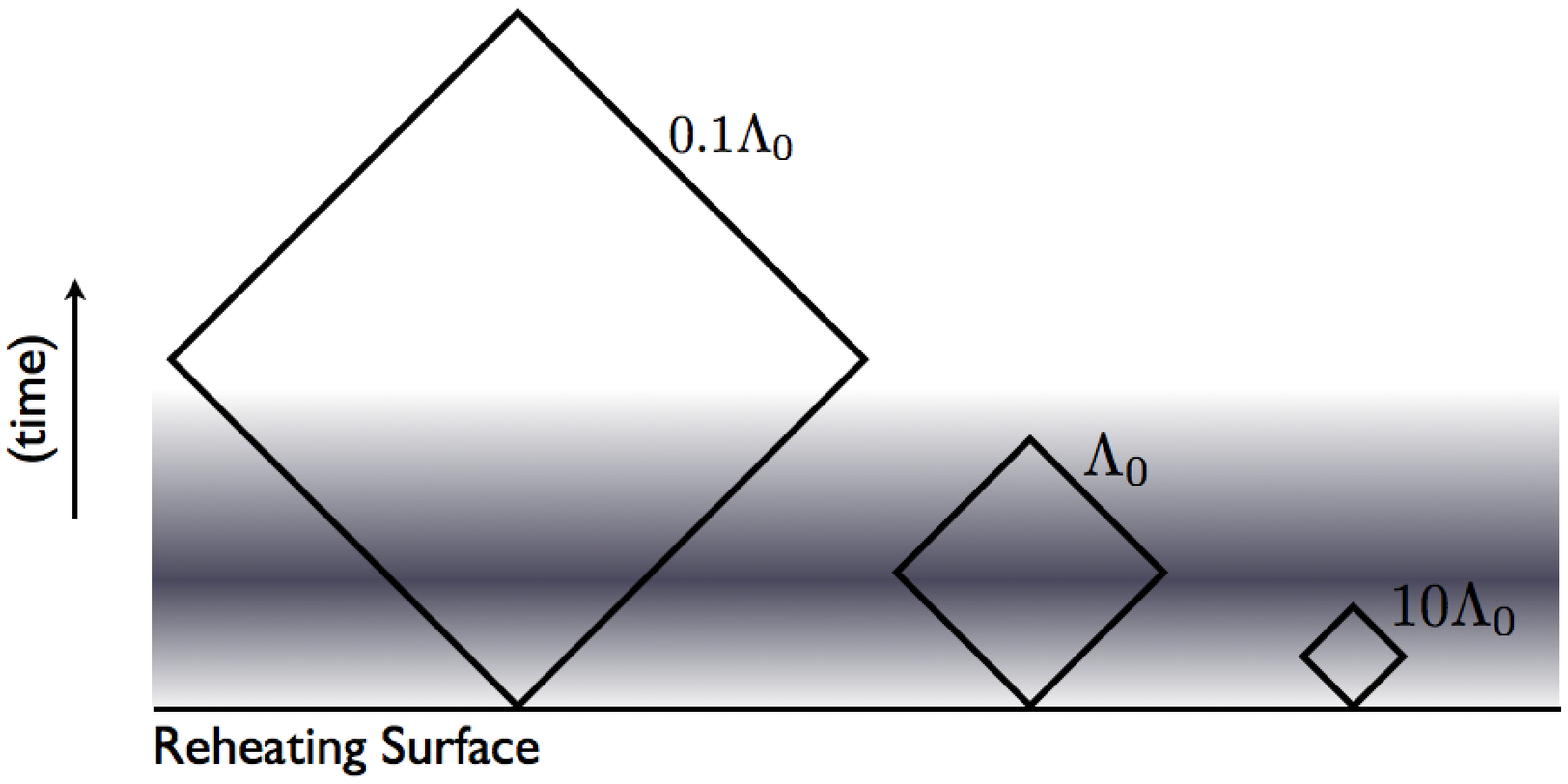,width=1.0\textwidth}{\label{fig-goldilocks}
  This cartoon demonstrates how the Causal Entropic Principle leads to
  a preferred value of the cosmological constant.  The horizontal band
  represents the rate of entropy production; darker areas correspond
  to a higher rate.  Vacua are weighted by $\Delta S$, the total
  amount of ``darkness'' inside a causal diamond.  Vacua with large
  cosmological constant are plentiful in the landscape, but they lead
  to small causal diamonds, which capture virtually no entropy
  production (right).  For some smaller value, the diamond will be
  just large enough to capture the bulk of the entropy production
  (center).  This is the preferred cosmological constant.  Larger
  diamonds may capture slightly more $\Delta S$ (left), but not in
  proportion to their size.  They correspond to vacua with very small
  cosmological constant which are much rarer in the landscape.
  Therefore they will be suppressed.}
Therefore, $dP/d\log \rho_\Lambda $ will be maximal for values of
$\rho_\Lambda$ such that
\begin{equation}
t_{\rm edge}(\rho_\Lambda)=t_{\rm peak}~.
\label{eq-fast}
\end{equation}
By Eqs.~(\ref{eq-tlambda}) and (\ref{eq-edge}), the observed value of
$\log\rho_\Lambda$ should be near
\begin{equation}
\log\rho_{\Lambda, {\rm peak}}\approx 
\log (0.006/t_{\rm peak}^2)\approx -123~.
\end{equation}
This rough estimate is borne out by our more careful calculation in
Sec.~\ref{sec-calculation}.  The excellent agreement of the observed
$\log\rho_\Lambda$, Eq.~(\ref{eq-cc}), with this prediction is
reflected in Fig.~\ref{fig-covol}, where it can be seen that the edge
time and the peak time really coincide for our universe.

The width of our distribution can also be understood in this manner.
Let $t_{\rm on}$ and $t_{\rm off}$ be the times at which the entropy
production rate is at half of its peak rate.  Using those values in
Eq.~(\ref{eq-fast}) gives roughly the $1\sigma$ boundaries we found
for our distribution in Sec.~\ref{sec-total}.  To summarize, the peak
and the width of the probability distribution for $\rho_\Lambda$ are
related to the peak and width of the entropy production rate by
Eq.~(\ref{eq-fast}).

Our distribution has a greater width than the distribution obtained
from the number of observers-per-baryon; this can be seen clearly in
Fig.~\ref{fig-showmethemoney}.  This is also not hard to understand.
In the traditional approach, nothing compensates for the exponential
growth of $dP/d\log \rho_\Lambda $ with $\log \rho_\Lambda$, until a
fairly sharp cutoff occurs when $\log \rho_\Lambda $ becomes large
enough to disrupt galaxy formation.  Hence, the preferred values of
$\log\rho_\Lambda $ are squeezed into a narrow interval, and the
observed value is strongly excluded.  In our approach, the spacetime
volume of the causal diamond depends inversely on $\rho_\Lambda$,
cancelling the pressure towards large values of $\rho_\Lambda$.  The
width of the probability curve is set only by the shape of the peak of
the entropy production rate (Fig.~6), which is fairly wide.

In this discussion we have pretended that $\rho_\Lambda$ does not
affect the entropy production rate.  In fact, this is an excellent
approximation.  In the vicinity the observed value of $\rho_\Lambda$,
the total entropy production depends on $\rho_\Lambda$ mainly through
the geometry of the causal diamond.  The probability density decreases
away from this maximum.  As a result, values of $\rho_\Lambda$ large
enough to disrupt galaxy formation are highly suppressed {\em even
  before we take into account the suppression of the entropy
  production rate resulting from this disruption}.

This points at another crucial difference between weighting by entropy
production in the causal diamond, and weighting by
observers-per-baryon: the preferred $\rho_\Lambda$ is set by
completely different physical processes, and hence, by essentially
unrelated timescales.  In the latter approach, one assumes that
observers require galaxies.  Then the disruption of galaxy formation
cuts off the exponential growth of $dP/d\log \rho_\Lambda$.  As a
result, the preferred $\log \rho_\Lambda$ is set by the time when
galaxies first form, and this gives a value that is too large compared
to Eq.~(\ref{eq-cc}).

In our approach, we do not assume that observers require galaxies.
The size of the causal diamond depends inversely on $\rho_\Lambda$,
allowing the preferred range of values for $\log\rho_\Lambda$ to be
set by the time-dependence of the entropy production rate.  The time
of peak entropy production by dust heated by starlight picks out the
value $\log \rho_\Lambda \approx -123$.  The time-scale when galaxies
form does not enter directly.  In our universe, the difference between
the two timescales amounts to ``only'' 3 orders of magnitude in the
preferred $\rho_\Lambda$, but it is easy to imagine other vacua in the
landscape where the era of peak entropy production is parametrically
separated from the era when galaxy halos become nonlinear (for
example, by a large galaxy cooling time).

\subsection{Statistical interpretation}
\label{sec-interpret}

It is worth emphasizing that it is entirely irrelevant whether the
observed $\rho_\Lambda$ is, say, $0.5\sigma$ above or $0.6\sigma$
below the median of our distribution.  We get to make only one
measurement.  There is no reason to expect this one data point to be
on the median (or on the peak) of the probability distribution.  But
we can expect that it will not be a very unlikely value.  Any value in
the $1\sigma$ region certainly qualifies as not unlikely.  The success
of the Causal Entropic Principle, its formal advantages aside, is not
that it predicts the precise value of $\rho_\Lambda$, but that our
distribution shows that the observed value was {\em not unlikely\/} to
have been observed.

Physicists have a great degree of confidence in certain theories that
make only statistical predictions, even though we are unable to make
more than a finite number of measurements, let alone test all the
consequences of a theory.  In this spirit, our result improves our
confidence in the Causal Entropic Principle and the underlying
landscape.  To improve our confidence further, we cannot repeat the
measurement of the cosmological constant, but we can extract other
predictions or postdictions and compare those to observation.

\acknowledgments We thank T.~Abel, E.~Baltz, L.~Bildsten, G.~Bothun, D.~Croton,
M.~Davis, L.~Hall, P.~Horava, C.~McKee, G.~Smoot, A.~Vilenkin,
M.~White, and especially E.~Quataert for helpful conversations.  We
are grateful to P.~Hopkins for providing more detail and numerical
values on quasar luminosities from Ref.~\cite{Hopkins:2006fq}, to
D.~Scott for an up-to-date plot of the grand unified photon spectrum, and A. Vilenkin for pointing out an error in the plots in earlier versions of this paper. We also thank Massimo Porrati for extensive discussions at the outset of this project.  RH, GDK, and GP thank the Aspen Center for Physics
where part of this work was completed.  The work was supported by the
Berkeley Center for Theoretical Physics, by a CAREER grant of the
National Science Foundation (RB), and by DOE grants DE-AC03-76SF00098
(RB), DE-AC02-76SF00515 (RH), and DE-FG02-96ER40969 (GDK).

\appendix

\section{The radiation era}
\label{sec-radiation}

In this Appendix, we justify our neglect of the radiation era.  The
metric, conformal time, and density during this era are
\begin{eqnarray} 
  a_{\rm rad}(t) &=& c (t-t_0)^{1/2}~, 
  \label{eq-arad}\\
  \tau_{\rm rad}(t) &=& 2c^{-1} (t-t_0)^{1/2}~, \\
  \rho_{\rm rad}(t) &= & \frac{3}{32\pi(t-t_0)^2}~.
  \label{eq-rhorad}
\end{eqnarray} 
The constants $c$ and $t_0$ are determined by matching the Hubble
constant and the scale factor to the metric Eq.~(\ref{eq-aexact}),
which becomes
\begin{equation}
a(t) = \left(\frac{3t}{2}\right)^{2/3}~
\end{equation}
for $t\ll t_\Lambda$.  They must agree at the time $ t_{\rm eq}$, when
the matter and radiation densities are equal, i.e.,
when~\cite{Tegmark:2005dy}
\begin{equation}
\rho_{\rm rad}=\rho_{\rm eq}\equiv 0.0026 \xi^4= 3.1\times 10^{-113}~, 
\end{equation}
where
\begin{equation}
\xi \approx 3.3 \times 10^{-28} 
\end{equation}
is the observed mass of pressureless matter per photon.  This yields
\begin{eqnarray}
  t_0 &=& \frac{t_{\rm eq}}{4} = 
  \frac{1}{6}\left(\frac{3}{8\pi\rho_{\rm eq}}\right)^{1/2}~,\\
  c &=& \left(\frac{24}{\pi\rho_{\rm eq}}\right)^{1/12}~.
\end{eqnarray}

By Eq.~(\ref{eq-cd}), the size of the causal diamond is set by the
total conformal time duration of the universe since reheating, which
is finite.  In Sec.~\ref{sec-metric}, we neglected the radiation era
and extended the matter/vacuum solution all the way back to the big
bang ($t=0$).  This yielded a total conformal time
\begin{equation}
\Delta\tau = 2.804 \left(\frac{3}{8\pi \rho_\Lambda}\right)^{1/6}~,
\end{equation}
from Eqs.~(\ref{eq-tlambda}) and (\ref{eq-tau0}).

In order to correct for the presence of the radiation era, we should
subtract the conformal time interval $\Delta\tau'$ corresponding to
the era $0<t<t_{\rm eq}$ that should be excised from the matter/vacuum
solution.  It should be replaced by the conformal time interval
$\Delta\tau''$ corresponding to the radiation dominated era (the
portion of the metric (\ref{eq-arad}) between reheating and matter
domination).  

Using the above results, however, it is easy to show
that
\begin{equation}
2\Delta\tau''<\Delta\tau' = 
\left(\frac{\rho_\Lambda}{\rho_{\rm eq}}\right)^{1/6}
\frac{\Delta\tau}{2.804}\,.
\end{equation}
Thus, the corrections to the conformal time, and thus to the size of
the causal diamond, are negligible for $\rho_\Lambda<\rho_{\rm eq}$.
For example, with the observed value of $\rho_\Lambda$, the correction
is less than 1\%.  The probability of values of
$\rho_\Lambda>10^{-120}$ almost vanishes according to our calculation;
yet this is still 7 orders of magnitude below $\rho_{\rm eq}$.  Hence,
our approximation is good in the entire range of $\rho_\Lambda$ in
which our probability distribution has support.

\bibliographystyle{JHEP}
\bibliography{all2}
\end{document}